\documentclass[11pt,a4paper]{article}

\pdfoutput=1
\usepackage[left=2.5cm,right=2.5cm,top=2.5cm,bottom=2.5cm]{geometry}
\usepackage[onehalfspacing]{setspace}
    \usepackage[pdftex,     
            plainpages=false,   
            breaklinks=true,    
            colorlinks=true
           ]{hyperref} 
\usepackage{thumbpdf}
\usepackage{amsmath}
\usepackage{amssymb}
\usepackage{amsfonts}
\usepackage{array}
\usepackage{bbold}
\usepackage{cancel}
\usepackage[hang,small]{caption2}
\usepackage{subfigure}
\usepackage{verbatim}
\usepackage{slashed}
\usepackage{graphicx}
\usepackage{multirow}
\usepackage{url}
\usepackage{cite}

\newcommand{\ii}{\mathrm{i}}
\newcommand{\rme}{\mathrm{e}}

\newcommand{\vev}[1]{{\left\langle #1 \right\rangle}}

\DeclareMathOperator{\tr}{tr}

\makeatletter
\newcommand*{\letterdef@}{}
\newcommand*{\letterdef}[3]{%
	\def\letterdef@##1{\expandafter\newcommand\csname #1\endcsname{#2{##1}}}%
	\@tfor\@tempa :=#3\do{\expandafter\letterdef@\expandafter{\@tempa}}}
\makeatother
\letterdef{c#1} {\mathcal}{ABCDEFGHIJKLMNOPQRSTUVWXYZ} 
\letterdef{rm#1}{\mathrm} {dDmM} 

\newcommand{\parenth}[1]{\left( #1 \right)}

\renewcommand{\a}{\alpha}

\numberwithin{equation}{section}

\newcommand{\be}{\begin{equation}}
\newcommand{\ee}{\end{equation}}

\makeatletter
\def\maketag@@@#1{\hbox{\m@th\normalfont\normalsize#1}}
\makeatother

\begin{document}

\begin{titlepage}

\vspace*{1cm}

\begin{center}{
\bfseries\LARGE
Correlators in superconformal quivers made QUICK}\\[8mm]
Michelangelo~Preti\footnote{E-mail: \texttt{michelangelo.preti@gmail.com}} \\[1mm]
\end{center}

\vspace*{0.50cm}

\centerline{\itshape
Mathematics Department, King’s College London}

\centerline{\itshape The Strand, London WC2R 2LS, UK}

\vspace*{1cm}
\begin{abstract}
In this paper we conclude the program of \cite{Galvagno:2020cgq,Galvagno:2021bbj} about perturbative approaches for $\mathcal{N}=2$ superconformal quiver theories in 4D.
We consider several classes of observables that involve multitrace local operators and Wilson loops scattered in all the possible ways among the quiver. We evaluate them exploiting the multi-matrix model arising from supersymmetric localisation and we generalise the solution to both $SU(N)$ and $U(N)$ cases. Moreover, we provide \texttt{QUICK} (\texttt{QUI}ver \texttt{C}orrelator \texttt{K}it) a Wolfram Mathematica package designed to automatise the perturbative solution of the $A_{q-1}$ multi-matrix model for all the observables mentioned above. Given the interpolating nature of the superconformal quiver theories $A_{q-1}$, the package is an efficient tool to compute correlators also in SCQCD, $\mathcal{N}=4$ SYM and its $\mathbb{Z}_q$ orbifolds.  
This manuscript includes a user guide and some pedagogical examples.  
\end{abstract}

\vspace{1cm}

\end{titlepage}

\section{Introduction}

The maximally supersymmetric theory in 4D ($\mathcal{N}=4$ SYM) plays a central role in the study of gauge theories. It provides
one of the most successful realisation of the AdS/CFT correspondence and it is the most favourable playground for obtaining exact results. Supersymmetric localisation is certainly one of the most effective technique to generate such results exploiting the BPS nature of the related observables. Those observables are, for instance, the 1/2 BPS circular Wilson loop and the 1/2 BPS chiral local operators. The expectation value of the circular Wilson loop was proven to localise to a Gaussian matrix model on a four sphere \cite{Erickson:2000af,Drukker:2000rr,Pestun:2007rz}, providing also the first non trivial test of the AdS/CFT correspondence. Moreover, supersymmetric localisation gives access to a richer class of observables such as general correlation functions that include local operators and Wilson loops  \cite{Semenoff:2001xp,Pestun:2002mr,Drukker:2007qr,Pestun:2009nn,Giombi:2009ds,Giombi:2009ek,Giombi:2012ep,Bonini:2014vta,Bonini:2015fng}, the Bremsstrahlung function \cite{Correa:2012at,Lewkowycz:2013laa}, multiple insertions of Wilson loops and Wilson loops in higher dimensional representations \cite{Okuyama:2018aij,Correa:2018lyl,Correa:2018pfn, CanazasGaray:2019mgq, Beccaria:2020ykg}. This technique was successfully applied to different frameworks as in 2D \cite{Panerai:2018ryw} and in the analogue of $\cN=4$ in 3D, namely $ABJM$ theory \cite{Kapustin:2009kz,Marino:2009jd,Drukker:2010nc,Bianchi:2018bke,Griguolo:2021rke} where several observables have been computed exactly \cite{Bianchi:2014laa,Bianchi:2017svd,Bianchi:2017ozk,Bianchi:2018scb,Drukker:2019bev}.

Sticking to the 4D case, the natural extension of the $\cN=4$ features is the less supersymmetric $\cN=2$ case. Despite the amount of supersymmetries is halved, localisation is powerful enough to reduce correlation function that includes chiral operators and Wilson loops to a matrix model. Unlike the $\cN=4$ SYM case, the resulting matrix model is no longer Gaussian. Among all the $\mathcal{N}=2$ theories, superconformal QCD (SCQCD) is the most studied. It is described by an $SU(N)$ gauge group with matter content given by $2N$ hypermultiplets. In the last few years, several observables were computed in $\mathcal{N}=2$ theories such as the Wilson loop  \cite{Andree:2010na,Passerini:2011fe,Bourgine:2011ie,Billo:2019fbi,Fiol:2020bhf,Beccaria:2021vuc}, its correlation function with chiral operators \cite{Rodriguez-Gomez:2016cem,Billo:2018oog,Beccaria:2020hgy}, the Bremsstrahlung function \cite{Fiol:2015mrp,Bianchi:2018zpb,Bianchi:2019dlw,Galvagno:2021qyq} and, recently, correlators at strong coupling \cite{Beccaria:2021hvt,Billo:2021rdb,Billo:2022xas,Beccaria:2022ypy}.

In this paper, we study a class of $\mathcal{N}=2$ theories with gauge structure given by a circular quiver with $q$ nodes and known as $A_{q-1}$. Those theories possess some peculiar properties. Indeed, under special conditions, they admit a holographic dual \cite{Kachru:1998ys,Gukov:1998kk} defined as a type IIB string theory on  AdS$_5 \times (S^5/\mathbb Z_q)$ orbifold, and they have been studied also from an integrability perspective \cite{Gadde:2009dj,Gadde:2010zi,Pomoni:2011jj,Gadde:2012rv,Pomoni:2013poa,Mitev:2014yba,Mitev:2015oty,Pomoni:2019oib,Pittelli:2019ceq,Niarchos:2019onf,Niarchos:2020nxk}. Moreover, they are known as "interpolating theories" since they are positioned between $\cN=4$ SYM and the conventional $\cN=2$ SCQCD. Supersymmetric localisation is very effective also in this context reducing BPS observables to a non-Gaussian multi-matrix model. Using this technique, recently several interesting results were obtained for Wilson loop vevs \cite{Fiol:2020ojn,Zarembo:2020tpf,Ouyang:2020hwd,Beccaria:2021ksw}, chiral/antichiral two- and three-point correlators \cite{Pini:2017ouj,Galvagno:2020cgq,Billo:2022gmq,Billo:2022fnb} and correlators of Wilson loops and local operators \cite{Galvagno:2021bbj}. Another interesting reason to study those theories in the perturbative regime is that they have a string realisation in the tensionless limit \cite{Gaberdiel:2022iot}.

In \cite{Galvagno:2020cgq, Galvagno:2021bbj} a complete weak coupling analysis of $A_{q-1}$ multi-matrix model has been developed. It was applied to the $SU(N)$ vev of Wilson loops, two-point correlators of chiral/anti-chiral multi-trace operators and general correlators of Wilson loops and a local operator. In this manuscript, we exploit the technical achievements obtained so far generalising the perturbative solution of the multi-matrix model also to the $U(N)$ case. Moreover, we provide the Mathematica package \texttt{QUICK} (\texttt{QUI}ver \texttt{C}orrelator \texttt{K}it) designed to automatise the perturbative solution of the $A_{q-1}$ multi-matrix model for all the observables mentioned above both for $SU(N)$ and $U(N)$ gauge groups. The package is built on 3 algorithms (one for any class of observables we consider) and it is also equipped by several options to explore all the possible configurations of the $A_{q-1}$ theories.
Once the number of nodes of the quiver $q$ is chosen, it generates the perturbative expansion up to the desired order in the couplings (or transcendental functions) of the selected observable. The only inputs needed are the dimension of the multi trace-operators and the vector multiplets in which operators and Wilson loops\footnote{Multiple coincident Wilson loops belonging to several vector multiplets can be considered.} belong. Given the features of the interpolating theory $A_{q-1}$, the \texttt{QUICK} package is extremely efficient to study those observables also in SCQCD, $\mathcal{N}=4$ SYM and its $\mathbb{Z}_q$ orbifolds. Attached to this manuscript, we provide an ancillary Mathematica notebook (\texttt{QUICKExample.nb}) with several examples and tutorials to use the package.

This paper is organised as follows. In section \ref{sec:setup} we introduce the $A_{q-1}$ theories and their properties. We also define the observables of interest both in the field theory and in the multi-matrix model picture providing a precise map between them. Moreover, we review the solution of the mixing problem in moving from the flat space to $S^4$ and we generalise the recursive solution of the multi-matrix model both for $SU(N)$ and $U(N)$. In section \ref{sec:algorithm} we define the 3 algorithms needed to solve the multi-matrix model together with examples. Finally, in section \ref{sec:QUICK} we show how the algorithms are implemented in the Mathematica package. We provide a detailed manual and several usage examples. Additional material is stored in the appendices.  

\section{Setup}\label{sec:setup}

\begin{figure}[!h]
\begin{center}
\includegraphics[scale=0.15,trim={15cm 2cm 15cm 2cm},clip]{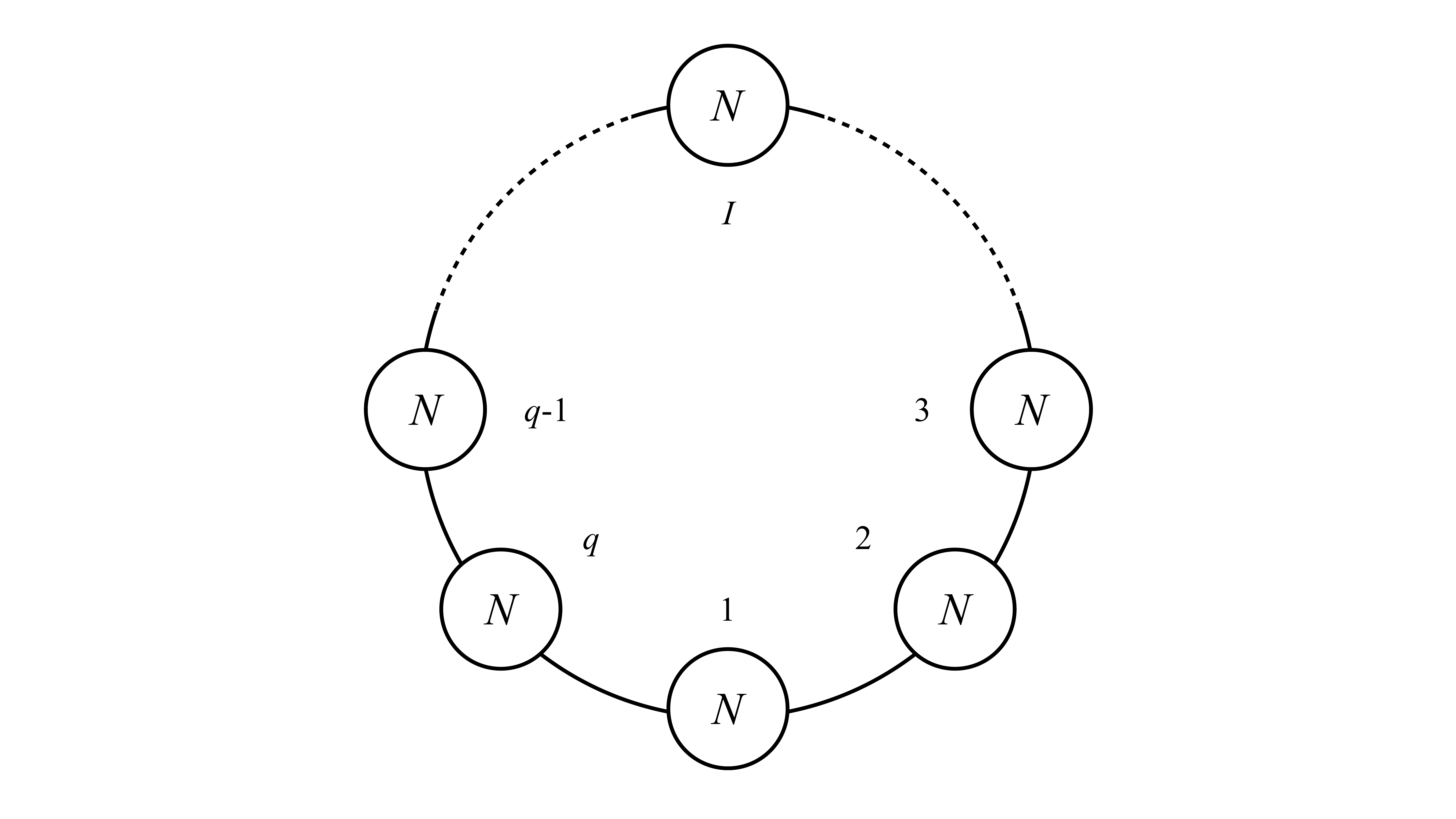}
\caption{$A_{q-1}$ theories as circular quivers with $q$ gauge nodes. Any node is labelled by an index $I=1,2,...,q$. $N$ is the rank of the gauge group $SU(N)$ or $U(N)$.}
\label{Fig:circquiver}
\end{center}
\end{figure}

\subsection{$\mathcal{N}=2$ superconformal quiver theories}
In this paper we consider the family of $\cN=2$ Lagrangian superconformal quiver theories known as $A_{q-1}$ recently studied 
in \cite{Galvagno:2020cgq,Galvagno:2021bbj}. 
Their gauge structure can be represented through the circular quiver diagram in figure \ref{Fig:circquiver}, where $q$ is the total number of nodes. Each node labelled by $I$ corresponds to a vector multiplet while each line going from the node $I$ to the node $I+1$ stands for a matter hypermultiplet so that the total action of the $A_{q-1}$ theory is given by
\begin{equation}\label{action}
    S_{q-1}=S_{\text{vector}}+S_{\text{hyper}}\,,
\end{equation}
where the gauge and matter actions are given in appendix \ref{sec:appendixA}. The field content of the theory can be written in a compact way using the $\cN=1$ formalism such that
\begin{align}\label{N2fields}
I\mathrm{-th}~ \mathrm{Vector}_{(\cN=2)} &= \big(V, \Phi \big)_I ~\quad\mathrm{adj~of~SU}(N)_I\quad\text{or}\quad \mathrm{U}(N)_I\notag \\
 \mathrm{Hyper}_{(\cN=2)} &= \big(Q, \widetilde{Q} \big)\quad\;\, \big(\square, \bar{\square}\big)  ~\mathrm{of~SU}(N)_I\!\times\! \mathrm{SU}(N)_{I+1}\quad\text{or}\quad\mathrm{U}(N)_I\!\times\! \mathrm{U}(N)_{I+1} ~,
\end{align}
where $V$ is a $\cN=1$ vector superfield and $\Phi,~Q,~\tilde{Q}$ are $\cN=1$ chiral superfields. Notice that unlike \cite{Galvagno:2020cgq,Galvagno:2021bbj}, here we consider both the $\mathrm{SU}(N)$ and $\mathrm{U}(N)$ cases. Finally, each vector multiplet action \eqref{Svector} brings a non-running coupling constant $g_I$, that can be rewritten in the usual 't Hooft combination as follows 
\begin{align}\label{tHooftcouplings}
{\color{blue}\lambda}_I = g_I^2 N~,
\end{align}
where $I=1,2,...,q$ is the label of the node of the quiver. We want to stress that in our investigation it is convenient to use the 't Hooft combination \eqref{tHooftcouplings} even if we will always keep $N$ finite.

One of the most interesting features of the $A_{q-1}$ theories is their role as interpolating theories between $\cN=2$ SCQCD and $\cN=4$ SYM. The first is obtained setting all the coupling constants to zero except one (${\color{blue}\lambda}_{I\neq 1} =0$) for $q=2$, the second in the limit in which all the couplings are equal  ${\color{blue}\lambda}_{1},...,{\color{blue}\lambda}_{q}={\color{blue}\lambda}$. The latter is also known as \textit{orbifold point} since reduces the $A_{q-1}$ theory to $q$ copies of $\mathcal{N}=4$ SYM, namely its $\mathbb Z_q$ orbifold. This has interesting consequences from a holographic perspective. Indeed, at the orbifold point $A_{q-1}$ theories do admit a dual geometry of the type AdS$_5 \times (S^5/\mathbb Z_q)$.

\subsubsection{1/2 BPS operators}\label{sec:WL&OinN2}
In this context, we introduce the 1/2 BPS multi-trace chiral local operators 
\begin{align}
	\label{Ondefinition}
		O^{(I)}_{\vec{n}}(x) \equiv 
\;C^{(I)}_{\vec{n}}\;\tr \varphi_I^{n_1}(x)\, \tr \varphi_I^{n_2}(x) \ldots\,\tr \varphi_I^{n_t}(x)~,
\end{align}
belonging to the $I$-th node of the quiver and labelled by the vector $\vec n=\{n_1,n_2,...,n_t\}$ with $t$ the number of traces. $C^{(I)}_{\vec{n}}$ is a normalisation constant.
The scalar field $\varphi_I$ is the one appearing in the vector multiplet \eqref{N2fields} and it corresponds to the first component of the $\Phi_I$ chiral superfield
\begin{equation}
\varphi_I(x) = \Phi_I(x,\theta,\bar \theta)\big|_{\theta = \bar \theta=0}~.
\end{equation}
The total R-charge of the operator \eqref{Ondefinition} is given by $n = \sum_{i=1}^t n_i$. Moreover, it is normal-ordered by construction.
Depending on the choice of the gauge group, the powers $n_i$ can take different values. Indeed, considering $SU(N)$, since $\tr \varphi_I=0$, one has $n_i\geq 2$, while choosing $U(N)$ also the value $n_i=1$ is allowed.

Another operator one can introduce in $A_{q-1}$ theories is the 1/2 BPS circular Wilson loop \cite{Rey:2010ry,Andree:2010na,Passerini:2011fe} which measures the holonomy of the gauge connection around a circular path $C$. Due to the presence of $q$ vector multiplets and then equivalently of $q$ gauge fields, in $A_{q-1}$ one can define $q$ Wilson loop operators, one for each node of the quiver as follows 
\begin{equation}
	\label{WLdef}
		W_I=\frac{1}{N}\tr \mathcal{P}
		\exp \bigg\{g_{I} \oint_C d\tau \Big[\ii \,A^{I}_{\mu}(x)\,\dot{x}^{\mu}(\tau)
		+\frac{R}{\sqrt{2}}\big(\varphi_{I}(x) +\bar \varphi_{I}(x)\big)\Big]\bigg\}\,,
\end{equation}
where $\{A^{I}_{\mu},\varphi_{I}\}$ are the gauge and scalar fields belonging to the $I$-th vector multiplet. The trace is taken over the fundamental of $SU(N)$ or $U(N)$ and  $x^\mu(\tau)$ parameterizes the circular path $C$ of radius $R$.

In this manuscript, we will consider observables built from the local and non-local operators defined above in \eqref{Ondefinition} and \eqref{WLdef}, that are captured by localisation.
This technique, based on supersymmetry, yields exact results for the following class of observables invariant with respect to a subset of the supersymmetry charges. 

The first observabel we consider is the vacuum expectation
value of multiple coincident Wilson loops\footnote{The circular loops have to be coincident to preserve enough supersymmetry to allow a localisation approach.}
\begin{equation}
\label{multWvev}
\vev{W_{\vec{I}}}_q\equiv\vev{W_{I_1}W_{I_2}...W_{I_n}}_q \equiv w_{\vec{I}}^{(q)}({\color{blue}\lambda}_1,...,{\color{blue}\lambda}_q,N)~,
\end{equation}
where $\vec{I}=[I_1,I_2,...,I_n]$ and $\vev{\;\;}_q$ represents the average computed in a theory with $q$ vector multiplets.
Each Wilson loop appearing in the left-hand side of \eqref{multWvev} can belong to any node of the quiver. When $n=1$, the vector $\vec{I}$ has only one entry $\vec{I}=[I_1]$, then the observable \eqref{multWvev} reduces to the expectation value of one Wilson loop belonging to the vector multiplet labelled by $I_1$.

The theory has enough supersymmetry to allows the localisation of some two-point functions. In particular we have the two-point function between chiral and anti-chiral operators defined as follows
\begin{equation}
	\label{twopointdef}
		\big\langle O^{(I)}_{\vec n}(x) \,\bar O^{(J)}_{\vec n}(0)\big\rangle_q 
		= \frac{\mathcal{G}^{(I,J)}_{\vec n}({\color{blue}\lambda}_1,...,{\color{blue}\lambda}_q,N)}{x^{2n}\phantom{\big|}}~,
\end{equation}
where the anti-chiral operator $\bar O^{(I)}_{\vec n}(x)$ is constructed as in \eqref{Ondefinition} but with the conjugate field $\bar \varphi_{I}(x)$. The form of the correlator \eqref{twopointdef} is fixed by (super-)conformal symmetry but, unlike the $\mathcal{N}=4$ SYM case, the coefficient $\mathcal{G}^{(I,J)}_{\vec n}$ is a non-trivial function of the couplings $\{\lambda_1,\lambda_2,...,\lambda_q\}$ and $N$. Furthermore, we have the two-point function between Wilson loops and a chiral local operator defined by
\begin{equation}
\label{OnWdef}
\vev{W_{\vec{I}}\;\,O^{(J)}_{\vec{n}}(0)}_q = \frac{\mathcal{A}_{\vec{n}}^{(\vec{I},J)}({\color{blue}\lambda}_1,...,{\color{blue}\lambda}_q,N)}{(2\pi R)^n}~.
\end{equation}
Since the circular Wilson loop can be interpreted as a superconformal defect, we can also refer to the correlation function \eqref{OnWdef} as the one-point function of the operator $O^{(J)}_{\vec{n}}$ in presence of the Wilson loops.

\subsection{The multi-matrix model}\label{sec:MM}

The partition function of $\cN=2$ Lagrangian theories localises to a finite dimensional integral on $S^4$ \cite{Pestun:2007rz}. In the present case of superconformal quiver theories $A_{q-1}$, it reduces to a multi-matrix model. The observables introduced in the previous section are invariant respect to the same supercharges that localises the partition function, then they are captured by taking suitable derivatives of the sphere partition function on the sphere, or equivalently by computing correlators in the
associated matrix model. In this section we review the localised partition function and the method to compute correlation functions in this framework as described in full details in \cite{Galvagno:2020cgq,Galvagno:2021bbj}.
The main feature of this procedure compared to the one of the eigenvalue distribution is its algorithmic structure. In section \ref{sec:QUICK} we present the package \texttt{QUICK} in which this method is implemented.

\subsubsection{From the localised partition function to correlators}

The partition function of the $A_{q-1}$ theories is given by the following multi-matrix model 
\begin{equation}
	\label{Zmulti}
		\cZ = \int \prod_{I=1}^q da_I ~ e^{-\tr a_I^2} \,\big| Z_{\mathrm{inst}}\big|^2\, \big| Z_{\mathrm{1-loop}} (a_I)\big|^2\,~,
\end{equation}
where to any node of the quiver labelled by $I$ it is associated a matrix $a_I$ that can be decomposed over the generators of $SU(N)$ or $U(N)$. Unlike the original proposal of \cite{Pestun:2007rz}, here we rescaled the matrices $a_I= \sqrt{g_{I}^2/(8\pi^2)} \;\tilde{a}_I$ to obtain a matrix model with normalised Gaussian factors and with a flat integration measure for each matrix defined as follows
\begin{align}
	\label{defMeasure}
 		da_I = \prod_{b} \frac{da_I^b}{\sqrt{2\pi}}~.
 \end{align}
Since in this paper we consider the perturbative sector of the matrix model \eqref{Zmulti}, the instanton partition function is set to $Z_{\mathrm{inst}}=1$. Besides, in the case in which one is interested only to the planar limit of the theory, the instantons contributions are exponentially suppressed anyway. The interaction terms of the matrix model originates from the 1-loop partition function $Z_{\mathrm{1-loop}}$ that reads
\begin{equation}
	\label{Z1loop}
	\big|Z_{\mathrm{1-loop}}\big|^2\,=\,\dfrac{\prod_{I} \prod_{i<j} H^2(a_i^I-a_j^I)}
	{\prod_{I} \prod_{i,j}^N H(a_i^I-a_j^{I+1})}
\qquad\text{with}\quad
\log H(x) = -\sum_{n=2}^\infty \frac{(-1)^n}{n} {\color{red}\zeta}_{2n-1} x^{2n}~,
\end{equation}
where ${\color{red}\zeta}_{2n-1}$ are the Riemann zeta functions ${\color{red}\zeta}({2n-1})$.

The perturbative approach to the matrix model \eqref{Zmulti} is based on the idea that the 1-loop determinant can be recast in the following exponential form 
\begin{equation}
	\label{ZtoSint}
	\big|Z_{\mathrm{1-loop}}\big|^2\, = \prod_{I=1}^q \mathrm{e}^{-\mathcal{S}_{\mathrm{int}}(a_I,a_{I+1})}~,
\end{equation}
where the exponent is interpreted as the interaction action given by
\begin{equation}\label{S_int}
\mathcal{S}_{\mathrm{int}} =  \!\!\sum_{m=2}^\infty \sum_{\ell=0}^{2m} \frac{(-1)^{m+\ell}}{(8\pi^2)^m m}\frac{{\color{blue}\lambda}_I^m}{ N^{2m}}\binom{2m}{\ell}  {{\color{red}\zeta}}_{2m-1} \!
\left[\tr a_I^{2m-\ell}\tr a_{I}^\ell-\frac{{\color{blue}\lambda}_{I+1}^{\ell/2}}{{\color{blue}\lambda}_I^{\ell/2}}\tr a_I^{2m-\ell}\tr a_{I+1}^\ell\right]~.
\end{equation} 
In \eqref{ZtoSint}, the product over $I$ takes into account the geometry of the quiver such that $a_{q+1}=a_1$. Thus, the partition function \eqref{Zmulti} takes the following form
\begin{equation}\begin{split}\label{partfun}
\cZ&=\int \prod_{I=1}^q \parenth{da_I~\rme^{-\tr\, a_I^2 - \cS_{\mathrm{int}}(a_I,a_{I+1})}} \\
&= \prod_{I=1}^q\vev{\rme^{- \cS_{\mathrm{int}}(a_I,a_{I+1})}}_0~,
\end{split}\end{equation}
where $\vev{\;\;}_0$ is the vev computed in the Gaussian model. 
Due to the rescaling of the matrices described above, the coupling constants are appearing only in $\cS_{\mathrm{int}}$. Then, computing \eqref{partfun} in perturbation theory at weak coupling corresponds to systematically expand $\rme^{- \cS_{\mathrm{int}}}$ in   ${\color{blue}\lambda}_I$ and treat the resulting terms as correlators in $q$ copies of the free Gaussian model. It is important to stress that the free Gaussian model, namely the theory with $\mathcal{S}_{\mathrm{int}}=0$, it corresponds precisely to $\cN=4$ SYM theory! 

Let's consider a simple example for the theory with 2 vector multiplets, namely $A_1$. Expanding $\mathcal{S}_{\mathrm{int}}$ at the first perturbative order, if one chooses $SU(N)$ as gauge group, \eqref{partfun} becomes
\begin{equation}\begin{split}\label{example11}
\cZ_{A_1} \!\!=&\!\! \!\int \!\!d a_1 d a_2\exp \!\!\Bigg[\!\!-\!\mathrm{tr}\, a_1^2\!-\!\mathrm{tr}\, a_2^2-\frac{3{\color{red}\zeta}_3}{64\pi^4} \Big[{\color{blue}g}_1^4(\mathrm{tr}\, a_1^2)^2\!+\!{\color{blue}g}_2^4(\mathrm{tr}\, a_2^2)^2\!\!-\!2{\color{blue}g}_1^2{\color{blue}g}_2^2\mathrm{tr}\, a_1^2\mathrm{tr}\, a_2^2\Big]\!\!+\!...\Bigg]\\
=&\vev{\mathbb{1}}_0\vev{\mathbb{1}}_0\!-\!\frac{3{\color{red}\zeta}_3}{64\pi^4}\!\Big[{\color{blue}g}_1^4\vev{(\mathrm{tr}\, a_1^2)^2}_0\vev{\mathbb{1}}_0 +{\color{blue}g}_2^4\vev{\mathbb{1}}_0\vev{(\mathrm{tr}\, a_2^2)^2}_0 \!-2{\color{blue}g}_1^2{\color{blue}g}_2^2\vev{\mathrm{tr}\, a_1^2}_0\vev{\mathrm{tr}\, a_2^2}_0\Big]\!+...
\end{split}\end{equation}
while for $U(N)$ it reads
\begin{equation}\begin{split}\label{example11UN}
\cZ_{A_1} =\vev{\mathbb{1}}_0\vev{\mathbb{1}}_0
\!-\!\frac{3{\color{red}\zeta}_3}{64\pi^4}\!\Big[
{\color{blue}g}_1^4[\vev{(\mathrm{tr}\, a_1^2)^2}_0-&\tfrac{4}{3}\vev{\mathrm{tr}\, a_1\mathrm{tr}\, a_1^3}_0]\vev{\mathbb{1}}_0
-2{\color{blue}g}_1^2{\color{blue}g}_2^2\vev{\mathrm{tr}\, a_1^2}_0\vev{\mathrm{tr}\, a_2^2}_0\\
&+{\color{blue}g}_2^4\vev{\mathbb{1}}_0[\vev{(\mathrm{tr}\, a_2^2)^2}_0-\tfrac{4}{3}\vev{\mathrm{tr}\, a_2\mathrm{tr}\, a_2^3}_0]
\Big]\!+...\,,
\end{split}\end{equation}
where $\mathbb{1}$ is the identity matrix and $\big\langle\mathbb{1}\big\rangle_0=1$. In the following, we will provide a recursive algorithm to compute all the correlators in the Gaussian model both for $SU(N)$ and $U(N)$.

This property can be extended from the partition function to any gauge invariant observable. Indeed, given an operator generically represented by the function $f(a_J^k)$, we have
\begin{align}
	\label{vevf}
		\vev{ f(a_J^k) }_q\, 
		\!=\! \frac{1}
		{\cZ} \int\! \prod_{I=1}^q da_I~\rme^{-\mathrm{tr}\, a_I^2 - \cS_{\mathrm{int}}(a_I,a_{I+1})}\,f( a_J^k)\,	\!=\! \frac{1}
		{\cZ} \,\prod_{I=1}^q\vev{
		\rme^{- \cS_{\mathrm{int}}(a_I,a_{I+1})}\,f( a_J^k)}_0~,
\end{align} 
where the expectation value of $f$ in the interacting matrix model is reduced to the computation of $\rme^{- \cS_{\mathrm{int}}}\,f$ in the free Gaussian model, namely $\cN=4$ SYM.

\paragraph{Recursion relations}$\\$

In order to compute the partition function \eqref{partfun} or the expectation value of an arbitrary operator $f$ \eqref{vevf}, one needs to study the multi-trace correlators in the Gaussian model $\vev{\tr a_I^{n_1}\tr a_I^{n_2}\dots}_0$ as explicitly shown in \eqref{example11}.
Then it is convenient to introduce the following notation
\begin{equation}
	\label{rectn}
		t^{(I)}_{[n_1,n_2,\dots]} = t^{(I)}_{\vec{n}}= \vev{\tr a_I^{n_1}\tr a_I^{n_2}\dots}_0~.
\end{equation}
The matrices $a_I$ can be written on a basis of $SU(N)$ or $U(N)$ generators $T_b$, with
$b = 1,\ldots, N^2-1$ or $b = 1,\ldots, N^2$ respectively, normalised as
\begin{equation}
\tr \,T_b \,T_c=\frac{1}{2}\,\delta_{bc}~,
\label{normtrace}
\end{equation}
writing each matrix as $a=a^b\,T_b$ with the "propagator" for the components given by $\vev{ a^b\, a^c}_0 = ~\delta^{bc}$.

Setting the following initial conditions
\begin{align}
	\label{rectnodd}
		t^{(I)}_0=N~,~~~~~~~\quad t^{(I)}_{\vec{n}} = 0~~~\text{for $\sum_i n_i$ odd}~,
\end{align}
any $t^{(I)}_{\vec{n}}$ defined in \eqref{rectn} can be evaluated solving the recursion relation originally derived in \cite{Billo:2017glv}. 
In the case of $SU(N)$ matrices, it reads
\begin{equation}\begin{split}\label{recursionSUN}
t^{(I)}_{[n_1,n_2,\dots,n_t]} =   &\frac{1}{2} \sum_{m=0}^{n_1-2}  \Big( t^{(I)}_{[m,n_1-m-2,n_2,\dots,n_t]}
		-\frac{1}{N}\,   t^{(I)}_{[n_1-2,n_2,\dots,n_t]}  \Big) \\ 
		&+ \sum_{k=2}^{t}\frac{n_k}{2} \,\Big(  t^{(I)}_{[n_1+n_k-2,n_2,\dots,\slashed{n_k},\dots,n_t]} -\frac{1}{N} \,t^{(I)}_{[n_1-1,n_2,\dots,n_k-1,\dots,n_t]} \Big)~,
\end{split}\end{equation}
while for $U(N)$ is reduced to
\begin{equation}\begin{split}\label{recursionUN}
t^{(I)}_{[n_1,n_2,\dots,n_t]} =   \frac{1}{2} \sum_{m=0}^{n_1-2}   t^{(I)}_{[m,n_1-m-2,n_2,\dots,n_t]}+ \sum_{k=2}^{t}\frac{n_k}{2}   t^{(I)}_{[n_1+n_k-2,n_2,\dots,\slashed{n_k},\dots,n_t]}~.
\end{split}\end{equation}
The notation ${[n_1,\dots,\slashed{n_k},\dots,n_t]}$ indicates the vector ${[n_1,\dots,n_t]}$ where the $k$-th indices is removed. One important difference between the two cases is the following. Since $\tr a_I=0$ if $a_I\in SU(N)$, any  $t^{(I)}_{\vec{n}}$ with at least an index $n_i=1$ is vanishing. On the other hand, if $a_I\in U(N)$, $t^{(I)}_{\vec{n}}$ is non-vanishing even if some indices are set to 1 as long as the sum of them is even (see \eqref{rectnodd}).

The recursive formulas \eqref{recursionSUN} and \eqref{recursionUN} originates from the fusion/fission identities satisfied by the $SU(N)$ and $U(N)$ generators respectively. The first is given by
\begin{equation}
\label{fussionSUN}
\begin{aligned}	
		\tr\big(T_b\, A\, T_b\, B\big) & = \frac{1}{2}\,\tr A\, \tr B
    	-\frac{1}{2N}\,\tr \big(A\, B\big)~,\\
		\tr \big(T_b\, A \big) \,\tr\big(T_b \,B\big) & 
		= \frac{1}{2}\,\tr\big(A\,B\big) -\frac{1}{2N}\,\tr A \,\tr B~,
\end{aligned}
\end{equation}
and the second reads
\begin{equation}
\label{fussionUN}
\begin{aligned}	
		\tr\big(T_b\, A\, T_b\, B\big) & = \frac{1}{2}\,\tr A\, \tr B~,\\
		\tr \big(T_b\, A \big) \,\tr\big(T_b \,B\big) & 
		= \frac{1}{2}\,\tr\big(A\,B\big) ~,
\end{aligned}
\end{equation}
where $A$ and $B$ are two arbitrary $N\times N$ matrices. In other words, the identities \eqref{fussionSUN} and \eqref{fussionUN} allows to relate any correlator $t^{(I)}_{\vec{n}}$ to a combination of other correlators obtained after a single Wick contraction. 

Let's consider for instance the correlators appearing at the first order in the partition function \eqref{example11} and \eqref{example11UN}. In the $SU(N)$ case we have 
\begin{equation}\label{texamplesun}
    t_{[2]}^{(1)}=t_{[2]}^{(2)}=\frac{N^2-1}{2}~,\qquad
    t_{[2,2]}^{(1)}=t_{[2,2]}^{(2)}=\frac{N^4-1}{4}~,
\end{equation}
while in the $U(N)$
\begin{equation}
    t_{[2]}^{(1)}=t_{[2]}^{(2)}=\frac{N^2}{2}~,\qquad
    t_{[2,2]}^{(1)}=t_{[2,2]}^{(2)}=\frac{N^2(N^2+2)}{4}~,\qquad
    t_{[1,3]}^{(1)}=t_{[1,3]}^{(2)}=\frac{3N^2}{4}~.
\end{equation}
Notice that, since any node of the superconformal quiver is associated to the same gauge group, for a given vector $\vec{n}$, the correlators $t^{(I)}_{\vec{n}}$ are all the same for any $I$. Plugging the values of the $t$-functions in \eqref{example11} and \eqref{example11UN} we obtain
\begin{equation}
    \begin{split}
\cZ_{A_1} &=1
-\frac{3{\color{red}\zeta}_3}{256\pi^4N^2}\!\Big[(N^4-1)({\color{blue}\lambda}_1^2+{\color{blue}\lambda}_2^2)-2(N^2-1)^2{\color{blue}\lambda}_1{\color{blue}\lambda}_2\Big]+...\qquad\text{for $SU(N)$}\\
\cZ_{A_1} &=1
-\frac{3{\color{red}\zeta}_3}{256\pi^4}\!\Big[(N^2-2)({\color{blue}\lambda}_1^2+{\color{blue}\lambda}_2^2)-2N^2{\color{blue}\lambda}_1{\color{blue}\lambda}_2\Big]+...\qquad\qquad\qquad\;\text{for $U(N)$}\,.
    \end{split}
\end{equation}

\subsubsection{Wilson loops and chiral operators in the multi-matrix model}\label{sec:W&OinMM}

The aim of this section is to define the matrix model counterpart of the local and non-local operators already introduced in section \ref{sec:WL&OinN2} and to identify the gauge theory observables with the matrix model ones.
First of all, the matrix model relative of the Wilson loop \eqref{WLdef} defined on a circle of radius $R=1$ is defined as follows \cite{Pestun:2007rz}
\begin{equation}
\label{Wlmm}
\mathcal{W}_I = \frac{1}{N}\tr\, \exp \left[\sqrt{\frac{{\color{blue}\lambda}_I}{2N}}\,a_I\right] = \frac{1}{N} \sum_{\ell = 0}^{\infty}\frac{1}{\ell!} \parenth{\frac{{\color{blue}\lambda}_I}{2 N}}^{\frac{\ell}{2}}\tr a_I^\ell~,
\end{equation}
then the operator corresponding to multiple coincident circular Wilson loops is given by
\begin{equation}
\label{Wlmultiplomm}
\mathcal{W}_{\vec{I}} \equiv
\mathcal{W}_{I_1}\mathcal{W}_{I_2}...\mathcal{W}_{I_n}=
\frac{1}{N^n} \sum_{\ell_1 ,\ell_2,...,\ell_n }\left[\prod_{i=1}^n\frac{1}{\ell_i!} \parenth{\frac{{\color{blue}\lambda}_{I_i}}{2 N}}^{\frac{\ell_i}{2}}\right]\tr a_{I_1}^{\ell_1}\tr a_{I_2}^{\ell_2}...\tr a_{I_n}^{\ell_n}~,
\end{equation}
The expectation value of one or several coincident Wilson loops in the matrix model coincides to the gauge theory one \eqref{multWvev}
\begin{equation}\label{WvecImm}
  w_{\vec{I}}^{(q)}({\color{blue}\lambda}_1,...,{\color{blue}\lambda}_q,N)
  =\vev{\mathcal{W}_{\vec{I}}}_q~.
\end{equation}
where the right-hand side is expressed in terms of the $t$-functions \eqref{rectn} that can be computed using the recursion relations \eqref{recursionSUN} or \eqref{recursionUN} depending on the choice of the gauge group.

The identification of the matrix model version of the multi-trace chiral operator introduced in \eqref{Ondefinition} is a non-trivial task. The most natural choice seems to be the replacement of the scalar fields in $O_{\vec{n}}^{(I)}$ with the matrices $a_I$ as follows 
\begin{align}
	\label{Onmmdefinition}
		\mathcal{O}^{(I)}_{\vec{n}} \equiv 
		C^{(I)}_{\vec{n}}\;\tr a_I^{n_1}\, \tr a_I^{n_2} \ldots\,\tr a_I^{n_t}~.
\end{align}
However, a fundamental property of the gauge theory operator is missing in \eqref{Onmmdefinition}. Indeed, since the gauge theory propagator involves a scalar field and its complex conjugate, the operator \eqref{Ondefinition} has no self-contraction by construction unlike \eqref{Onmmdefinition}. Therefore, we need to impose the cancellation of all the self-contraction from $\mathcal{O}^{(I)}_{\vec{n}}$ making it normal-ordered \cite{Gerchkovitz:2016gxx,Rodriguez-Gomez:2016cem,Rodriguez-Gomez:2016ijh,Billo:2017glv,Billo:2018oog}. In the following we review this procedure for $A_{q-1}$ theories described in depth in  \cite{Galvagno:2020cgq, Galvagno:2021bbj}.

Let's consider an operator $\mathcal{O}^{(I)}_{\vec n}$ with scaling dimension $n$. Its normal-ordered counterpart is given by the operator itself plus the linear combination of all the operators with dimensions $\{p\}=\{n-2,n-4,...\}$. The dimensions appearing in $\{p\}$ differ by 2 since we trade two matrices with a self-contraction. Unlike the case in which $n$ is odd, when it is even the last scaling dimension appearing in the set $\{p\}$ is 0 corresponding with the identity operator. For this reason, in order to threat alike the even and odd $n$ cases, one can define a slightly modified version of \eqref{Onmmdefinition} as follows
\begin{equation}\label{O-id}
    \tilde{\mathcal{O}}^{(I)}_{\vec n}\equiv
    \mathcal{O}^{(I)}_{\vec n}-\vev{\mathcal{O}^{(I)}_{\vec n}}_q\,,
\end{equation}
where we subtract the identity operator contribution.

Then, the normal-ordered operator of scaling dimension $n$ is given by
\begin{equation}\label{basis}
:\mathcal{O}^{(I)}_{\vec n}:=\tilde{\mathcal{O}}^{(I)}_{\vec n}+\sum_{\substack{\vec p=\text{partitions} \\ \text{of dim. $\{p\}$}}}
\;\sum_{\substack{J=\text{nodes of} \\ \text{quiver $A_{q-1}$}}}
\alpha^{(I,J)}_{\vec n,\vec p}\;\tilde{\mathcal{O}}^{(J)}_{\vec p}~,
\end{equation}
with the coefficients $\alpha=\alpha({\color{blue}\lambda}_1,...,{\color{blue}\lambda}_q,N)$.
Notice that, given the definition \eqref{O-id}, the basis of operators $\{\tilde{\mathcal{O}}^{(J)}_{\vec p}\}$ starts to contribute for $n\geq 3$ since in the other cases $\{p\}=\varnothing$.
The sum over $J$ is required to include all the operators belonging to any node of the quiver and the sum over $\vec{p}$ takes into account all the possible multi-trace operators with a given scaling dimension appearing in the set $\{p\}$. We want to stress that the former sum is highly dependent on the choice of the gauge group. Indeed, while the partitions $\vec{p}$ and the set $\{p\}$ can contain 1 if the gauge group is $U(N)$, they cannot in the $SU(N)$ case (see table \ref{tab:table1}).

\begin{table}[!t]
\begin{center}
 \begin{tabular}{||c | c c||} 
 \hline
 \multirow{2}{*}{dimension $n$} & 
 \multicolumn{2}{c||}{Operators $\in\,\{\tilde{\mathcal{O}}^{(J)}_{\vec p}\}$} \\
 \cline{2-3}
 & $SU(N)$ & $U(N)$ \\ 
 \hline\hline
 3 & $\varnothing$ & $\{\tilde{\mathcal{O}}^{(J)}_{[1]}\}$  \\ 
 \hline
 4 & $\{\tilde{\mathcal{O}}^{(J)}_{[2]}\}$ & $\{\tilde{\mathcal{O}}^{(J)}_{[2]},\tilde{\mathcal{O}}^{(J)}_{[1,1]}\}$ \\
 \hline
 5 & $\{\tilde{\mathcal{O}}^{(J)}_{[3]}\}$ & $\{\tilde{\mathcal{O}}^{(J)}_{[3]},\tilde{\mathcal{O}}^{(J)}_{[2,1]},\tilde{\mathcal{O}}^{(J)}_{[1,1,1]},\tilde{\mathcal{O}}^{(J)}_{[1]}\}$  \\
 \hline
 6 & $\{\tilde{\mathcal{O}}^{(J)}_{[4]},\tilde{\mathcal{O}}^{(J)}_{[2,2]},\tilde{\mathcal{O}}^{(J)}_{[2]}\}$ & $\{\tilde{\mathcal{O}}^{(J)}_{[4]},\tilde{\mathcal{O}}^{(J)}_{[3,1]},\tilde{\mathcal{O}}^{(J)}_{[2,2]},\tilde{\mathcal{O}}^{(J)}_{[2,1,1]},\tilde{\mathcal{O}}^{(J)}_{[1,1,1,1]},\tilde{\mathcal{O}}^{(J)}_{[2]},\tilde{\mathcal{O}}^{(J)}_{[1,1]}\}$  \\
 \hline
\end{tabular}
\end{center}
\caption{The basis of operators appearing in the normal ordering of the first few operators of dimension $n$ both for $SU(N)$ and $U(N)$ gauge groups.}
\label{tab:table1}
\end{table}

The coefficients appearing in \eqref{basis} can be determined imposing the orthogonality of the operators of the basis, namely through the Gram-Schmidt procedure.
Let's consider $\{\tilde{\mathcal{O}}^{(J)}_{\vec p}\}$ and $\{\tilde{\mathcal{O}}^{(K)}_{\vec s}\}$ two copies of the basis appearing in the linear combination of  $:\mathcal{O}^{(I)}_{\vec n}:$. The matrix of their mixed correlators reads
\begin{equation}\label{matrixM}
M_{\vec s,\vec p}^{(K,J)}=\langle \tilde{\mathcal{O}}^{(K)}_{\vec s} \;\tilde{\mathcal{O}}^{(J)}_{\vec p} \rangle_q\,,
\end{equation}
with $K,J=1,2,...,q$. Then the Gram-Schmidt coefficients $\alpha$ are given by
\begin{equation}\label{coeffGS}
\alpha^{(I,J)}_{\vec n,\vec p}=-\sum_{\substack{\text{nodes $K$} \\ \text{partitions $\vec s$ of $\{s\}$}}}\langle \tilde{\mathcal{O}}^{(I)}_{\vec n} \;\tilde{\mathcal{O}}^{(K)}_{\vec s} \rangle_q\;\left(M_{\vec s,\vec p}^{(K,J)}\right)^{-1}~,
\end{equation}
where $M^{-1}$ is the inverse of the matrix \eqref{matrixM}.

Finally, similarly to its gauge theory relative, the matrix model normal-ordered operator \eqref{basis} with coefficients given by \eqref{coeffGS} is orthogonal to all the lower-dimensional operators and consequently it doesn't present any self-contractions. In particular its one-point function vanishes
\begin{equation}
\langle \;:\mathcal{O}^{(I)}_{\vec n}: \;\rangle_q=0~.
\end{equation}
Since the equivalence between the gauge theory and the matrix model operators is established, the two-point function \eqref{twopointdef} in terms of the matrix-model operators becomes
\begin{equation}
	\label{Gnormord}
		\mathcal{G}^{(I,J)}_{\vec n}({\color{blue}\lambda}_1,...,{\color{blue}\lambda}_q,N)=\big\langle :\mathcal{O}^{(I)}_{\vec n}: \;:\mathcal{O}^{(J)}_{\vec n}:\big\rangle_q ~,
\end{equation}
and equivalently the one-point function in presence of Wilson loops \eqref{OnWdef} reads
\begin{equation}\label{Adef}
    \mathcal{A}_{\vec{n}}^{(\vec{I},J)}({\color{blue}\lambda}_1,...,{\color{blue}\lambda}_q,N)=\vev{\mathcal{W}_{\vec{I}}\;\,:\mathcal{O}^{(J)}_{\vec{n}}:}_q~.
\end{equation}
To make the this map consistent, we have to choose the normalisation of the local operator following the definition \eqref{Onmmdefinition}. Normal ordered operators appearing in the \eqref{Gnormord} have normalisation ${C}^{(I)}_{\vec{n}}=\frac{\lambda_I^{n/2}}{(8\pi^2N)^{n/2}}$ while the one appearing in \eqref{Adef} has normalisation ${C}^{(I)}_{\vec{n}}=(\lambda_I/N)^{\tfrac{n-2t}{2}}$ with $n=\sum n_i$ and $t$ the number of traces of the operator.

\section{Algorithms for correlators}\label{sec:algorithm}

In this section we provide 3 algorithms to compute the observables $\mathcal{G}$, $w$ and $\mathcal{A}$ defined in the previous section in \eqref{Gnormord}, \eqref{WvecImm} and \eqref{Adef}.

\paragraph{Algorithm 1: Correlators of two local operators $\mathcal{G}^{(I,J)}_{\vec n}$}

\begin{enumerate}
    \item Identify the basis of operators $\{\tilde{\mathcal{O}}^{(J)}_{\vec p}\}$ appearing in the normal-ordering of $\mathcal{O}^{(I)}_{\vec n}$ as in \eqref{basis}.
    \item Expand $\big\langle :\mathcal{O}^{(I)}_{\vec n}: \;:\mathcal{O}^{(J)}_{\vec n}:\big\rangle_q$ in terms of the matrix model correlators and Gram-Schmidt coefficients.
    \item Re-write the Gram-Schmidt coefficients appearing in the expansion above in terms of matrix model correlators as in \eqref{coeffGS}. 
    \item Express all the matrix model correlators on the sphere in terms of the t-functions \eqref{rectn} by expanding $\mathcal{S}_{\text{int}}$ in \eqref{vevf} to the relevant order in the couplings.
    \item Compute the resulting t-functions by means of the recursion relation \eqref{recursionSUN} or \eqref{recursionUN} depending on the choice of the gauge group. 
\end{enumerate} 

For example, let's consider the correlator $\mathcal{G}^{(1,2)}_{[4]}=\big\langle :\mathcal{O}^{(1)}_{[4]}: \;:\mathcal{O}^{(2)}_{[4]}:\big\rangle_2$ in $SU(N)$. As shown in Table \ref{tab:table1}, the normal-ordered operator $:\mathcal{O}_{[4]}^{(I)}:$ contains only operators of dimension 2 namely
\begin{equation}\begin{split}\label{O4NO}
    :\mathcal{O}_{[4]}^{(1)}:&=\tilde{\mathcal{O}}_{[4]}^{(1)}+\alpha_{[4],[2]}^{(1,1)}\;\tilde{\mathcal{O}}_{[2]}^{(1)}+\alpha_{[4],[2]}^{(1,2)}\;\tilde{\mathcal{O}}_{[2]}^{(2)}\,,\\
    :\mathcal{O}_{[4]}^{(2)}:&=\tilde{\mathcal{O}}_{[4]}^{(2)}+\alpha_{[4],[2]}^{(2,1)}\;\tilde{\mathcal{O}}_{[2]}^{(1)}+\alpha_{[4],[2]}^{(2,2)}\;\tilde{\mathcal{O}}_{[2]}^{(2)}\,.
\end{split}
\end{equation}
Substituting \eqref{O4NO} into the $\mathcal{G}^{(1,2)}_{[4]}$ definition, one can express the field theory correlator in terms of the matrix model two-point functions and the Gram-Schmidt coefficients as follows
\begin{equation}\begin{split}\label{G4}
\mathcal{G}^{(1,2)}_{[4]}=
\big\langle \tilde{\mathcal{O}}_{[4]}^{(1)}\tilde{\mathcal{O}}_{[4]}^{(2)}\big\rangle_2&+
\sum_{K=1}^2\left[\alpha_{[4],[2]}^{(1,K)}\big\langle \tilde{\mathcal{O}}_{[2]}^{(K)}\tilde{\mathcal{O}}_{[4]}^{(2)}\big\rangle_2+\alpha_{[4],[2]}^{(2,K)}\big\langle \tilde{\mathcal{O}}_{[4]}^{(1)}\tilde{\mathcal{O}}_{[2]}^{(K)}\big\rangle_2\right]\\
&+\sum_{K=1}^2\sum_{L=1}^2\alpha_{[4],[2]}^{(1,K)}\alpha_{[4],[2]}^{(2,L)}\big\langle \tilde{\mathcal{O}}_{[2]}^{(K)}\tilde{\mathcal{O}}_{[2]}^{(L)}\big\rangle_2\,.
\end{split}  
\end{equation}
The next step is to write the Gram-Schmidt coefficients in terms of combinations of the mixed correlators as in \eqref{coeffGS}. In this simple example, $M$ defined in \eqref{matrixM} is a 2x2 matrix and its inverse is given by
\begin{equation}
    \left(M_{[2],[2]}^{(K,J)}\right)^{-1}=\frac{1}{\big\langle \tilde{\mathcal{O}}_{[2]}^{(1)}\tilde{\mathcal{O}}_{[2]}^{(1)}\big\rangle_2\big\langle \tilde{\mathcal{O}}_{[2]}^{(2)}\tilde{\mathcal{O}}_{[2]}^{(2)}\big\rangle_2-\big\langle \tilde{\mathcal{O}}_{[2]}^{(1)}\tilde{\mathcal{O}}_{[2]}^{(2)}\big\rangle_2^2}\begin{pmatrix}
\big\langle \tilde{\mathcal{O}}_{[2]}^{(2)}\tilde{\mathcal{O}}_{[2]}^{(2)}\big\rangle_2 & -\big\langle \tilde{\mathcal{O}}_{[2]}^{(1)}\tilde{\mathcal{O}}_{[2]}^{(2)}\big\rangle_2 \\
-\big\langle \tilde{\mathcal{O}}_{[2]}^{(1)}\tilde{\mathcal{O}}_{[2]}^{(2)}\big\rangle_2 & \big\langle \tilde{\mathcal{O}}_{[2]}^{(1)}\tilde{\mathcal{O}}_{[2]}^{(1)}\big\rangle_2
\end{pmatrix}.
\end{equation}
Then the $\alpha$'s appearing in the normal ordered operator $:\mathcal{O}_{[4]}^{(1)}:$ in \eqref{O4NO} are the following
\begin{equation}\begin{split}\label{alphatest}
 \alpha_{[4],[2]}^{(1,1)}=&
\frac{\big\langle \tilde{\mathcal{O}}_{[4]}^{(1)}\tilde{\mathcal{O}}_{[2]}^{(2)}\big\rangle_2\big\langle \tilde{\mathcal{O}}_{[2]}^{(1)}\tilde{\mathcal{O}}_{[2]}^{(2)}\big\rangle_2-\big\langle \tilde{\mathcal{O}}_{[4]}^{(1)}\tilde{\mathcal{O}}_{[2]}^{(1)}\big\rangle_2\big\langle \tilde{\mathcal{O}}_{[2]}^{(2)}\tilde{\mathcal{O}}_{[2]}^{(2)}\big\rangle_2}{\big\langle \tilde{\mathcal{O}}_{[2]}^{(1)}\tilde{\mathcal{O}}_{[2]}^{(1)}\big\rangle_2\big\langle \tilde{\mathcal{O}}_{[2]}^{(2)}\tilde{\mathcal{O}}_{[2]}^{(2)}\big\rangle_2-\big\langle \tilde{\mathcal{O}}_{[2]}^{(1)}\tilde{\mathcal{O}}_{[2]}^{(2)}\big\rangle_2^2}\,,\\
\alpha_{[4],[2]}^{(1,2)}=&
\frac{\big\langle \tilde{\mathcal{O}}_{[4]}^{(1)}\tilde{\mathcal{O}}_{[2]}^{(1)}\big\rangle_2\big\langle \tilde{\mathcal{O}}_{[2]}^{(1)}\tilde{\mathcal{O}}_{[2]}^{(2)}\big\rangle_2-\big\langle \tilde{\mathcal{O}}_{[4]}^{(1)}\tilde{\mathcal{O}}_{[2]}^{(2)}\big\rangle_2\big\langle \tilde{\mathcal{O}}_{[2]}^{(1)}\tilde{\mathcal{O}}_{[2]}^{(1)}\big\rangle_2}{\big\langle \tilde{\mathcal{O}}_{[2]}^{(1)}\tilde{\mathcal{O}}_{[2]}^{(1)}\big\rangle_2\big\langle \tilde{\mathcal{O}}_{[2]}^{(2)}\tilde{\mathcal{O}}_{[2]}^{(2)}\big\rangle_2-\big\langle \tilde{\mathcal{O}}_{[2]}^{(1)}\tilde{\mathcal{O}}_{[2]}^{(2)}\big\rangle_2^2}\,,  
\end{split}
\end{equation}
while the coefficients appearing in  $:\mathcal{O}_{[4]}^{(2)}:$ are the same exchanging the node indices $1\leftrightarrow 2$. Plugging them into \eqref{G4}, one obtains $\mathcal{G}^{(1,2)}_{[4]}$ written in terms of matrix model correlators only. According to the point 4. of the algorithm, they can be expressed in terms of the t-functions, for instance  
\begin{equation}
\big\langle \tilde{\mathcal{O}}_{[2]}^{(1)}\tilde{\mathcal{O}}_{[2]}^{(2)}\big\rangle_2=
\frac{3}{32\pi^4}{\color{red}\zeta}_3
{\color{blue}\lambda}_1{\color{blue}\lambda}_2
(t^{(1)}_{[2]}t^{(1)}_{[2]}-t^{(1)}_{[2,2]})(t^{(2)}_{[2]}t^{(2)}_{[2]}-t^{(2)}_{[2,2]})+...\,,
\end{equation}
where $...$ stand for higher orders in the coupling constants. Finally solving the recursion relation \eqref{recursionSUN} for all the t's appearing in the expansion, as for instance in \eqref{texamplesun}, one obtains
\begin{equation}\label{sphereOOtest}
\big\langle \tilde{\mathcal{O}}_{[2]}^{(1)}\tilde{\mathcal{O}}_{[2]}^{(2)}\big\rangle_2=
\frac{3(N^2-1)^2{\color{red}\zeta}_3
{\color{blue}\lambda}_1{\color{blue}\lambda}_2}{128\pi^4N^2}
-\frac{5(N^2-1)^2(2N^2-3){\color{red}\zeta}_5
{\color{blue}\lambda}_1{\color{blue}\lambda}_2({\color{blue}\lambda}_1+{\color{blue}\lambda}_2)}{1024\pi^6N^4}
+...\,.
\end{equation}
Repeating the same procedure for all the matrix model two-point functions appearing in $\mathcal{G}^{(1,2)}_{[4]}$ and plugging all together we have
\begin{equation}\label{G124}
\mathcal{G}^{(1,2)}_{[4]}=
\frac{16(N^2-1)^2
{\color{blue}\lambda}_1^4{\color{blue}\lambda}_2^4\left(
72(2N^3-3N)^2{\color{red}\zeta}_3^2+35(N^4-6N^2+18)^2{\color{red}\zeta}_7\right)
}{(16\pi^2)^8N^{12}}
+...\,.
\end{equation}

\paragraph{Algorithm 2: Expectation value of Wilson loops $w^{(q)}_{\vec I}$}

\begin{enumerate}
    \item Use the definition of Wilson loop \eqref{Wlmm} or \eqref{Wlmultiplomm} in terms of the matrices $a_I$.
    \item Expand the action $\mathcal{S}_{\text{int}}$ in \eqref{vevf} and re-write the result in terms of the t-functions \eqref{rectn}.
\end{enumerate}
\underline{Option a}: $w^{(q)}_{\vec I}$ as a series in the couplings ${\color{blue}\lambda}_I$
\begin{enumerate}
\setcounter{enumi}{2}
    \item Evaluate the sums appearing in \eqref{Wlmm} or \eqref{Wlmultiplomm} up to suitable cutoffs and expand to the relevant order in the couplings.  
    \item Compute the resulting t-functions by means of the recursion relation \eqref{recursionSUN} or \eqref{recursionUN} depending on the choice of the gauge group.
\end{enumerate}
\underline{Option b}: $w^{(q)}_{\vec I}$ as a series in transcendental functions  ${\color{red}\zeta}_n$
\begin{enumerate}
\setcounter{enumi}{2}
    \item Use the recursion relations \eqref{recursionSUN} or \eqref{recursionUN} until the resulting t-functions $t^{(I)}_{\vec n}$ have all the elements of the vector $\vec n$ depending on combinations of the indices of the sums $\ell_i$.
    \item Perform a suitable shift on the indices $\ell_i$ to reduce the result to a combination of the t-functions appearing in the definitions \eqref{Wlmm} and \eqref{Wlmultiplomm} multiplied by some polynomial in $\ell_i$.
    \item Solve the sums in $\ell_i$ using the definitions \eqref{Wlmm} and \eqref{Wlmultiplomm} and the
    archetypal formula
    \begin{equation}\label{derdef}
    \frac{1}{N^n} \sum_{\ell_1 = 0}^{\infty}...\sum_{\ell_n = 0}^{\infty}\frac{(\ell_1+...+\ell_n)^k}{\ell_1!...\ell_n!} \parenth{\frac{{\color{blue}\lambda}_J}{2 N}}^{\frac{\ell_1+...+\ell_n}{2}}\;t_{[\ell_1,...,\ell_n]}=2^k [{\color{blue}\lambda}_J\partial_{J}]^k 
    w_{\vec{I}}~,
    \end{equation}
    where $\partial_{J}X=dX/d{\color{blue}\lambda}_J$ and $k$ is the number of nested applications of the differential operator to $w_{\vec{I}}\equiv \langle\mathcal{W}_{\vec{I}}\rangle_0$. The latter is the expectation value of the Wilson loops in the pure Gaussian model namely $q$ copies of $\mathcal{N}=4$ SYM. For instance, for a single Wilson loop in the node $I$ of the quiver we have \cite{Erickson:2000af,Drukker:2000rr,Pestun:2007rz}
    \begin{equation}
        \begin{split}\label{wIexactN=4}
            w_I({\color{blue}\lambda}_I,N)=& \frac{1}{N}\,L_{N-1}^{1}\Big(-\frac{{\color{blue}\lambda}_I}{4 N}\Big)\,\exp\left[\frac{{\color{blue}\lambda}_I}{8N}\Big(1-\frac{1}{N}\Big)\right]\qquad \text{for } SU(N)\\
            w_I({\color{blue}\lambda}_I,N)=& \frac{1}{N}\,L_{N-1}^{1}\Big(-\frac{{\color{blue}\lambda}_I}{4 N}\Big)\,\exp\left[\frac{{\color{blue}\lambda}_I}{8N}\right]\qquad\qquad\qquad \text{for } U(N)
        \end{split}
    \end{equation}
    or for two coincident Wilson loops lying on the same node of the quiver we have \cite{Drukker:2000rr,Kawamoto:2008gp,Okuyama:2018aij}
    \begin{equation}\small\begin{split}\label{wIIexactN=4}
    &w_{[I,I]}= \frac{e^{\frac{{\color{blue}\lambda}_I}{2N}\left(1-\frac{1}{N}\right)}}{N^2}L_{N-1}^{1}\Big(\frac{-{\color{blue}\lambda}_I}{N}\Big)\\
    &\quad+\frac{2e^{\frac{{\color{blue}\lambda}_I}{4N}\left(1-\frac{2}{N}\right)}}{N^2}\sum_{i=0}^{N-1}\sum_{j=0}^{i-1}\!\left[
    L_{i}\Big(\frac{-{\color{blue}\lambda}_I}{4N}\Big)
    L_{j}\Big(\frac{-{\color{blue}\lambda}_I}{4N}\Big)\!-
    \frac{j!}{i!}\Big(\frac{{\color{blue}\lambda}_I}{4N}\Big)^{i-j}
    L_{j}^{i-j}\Big(\frac{-{\color{blue}\lambda}_I}{4N}\Big)^2\right]\quad\!\! \text{for } SU(N)\\
    &w_{[I,I]}= \frac{e^{\frac{{\color{blue}\lambda}_I}{2N}}}{N^2}L_{N-1}^{1}\Big(\frac{-{\color{blue}\lambda}_I}{N}\Big)\\
    &\quad+\frac{2e^{\frac{{\color{blue}\lambda}_I}{4N}}}{N^2}\sum_{i=0}^{N-1}\sum_{j=0}^{i-1}\!\left[
    L_{i}\Big(\frac{-{\color{blue}\lambda}_I}{4N}\Big)
    L_{j}\Big(\frac{-{\color{blue}\lambda}_I}{4N}\Big)\!-
    \frac{j!}{i!}\Big(\frac{{\color{blue}\lambda}_I}{4N}\Big)^{i-j}
    L_{j}^{i-j}\Big(\frac{-{\color{blue}\lambda}_I}{4N}\Big)^2\right]\quad\qquad\; \text{for } U(N)
    \end{split}\end{equation}
    and so on. In \eqref{wIexactN=4} and \eqref{wIIexactN=4}, $L_n^k$ is the generalised Laguerre polynomial with $L_i=L_i^0$. 
\end{enumerate}

As an example, let's consider the expectation value $w_{[1]}^{(2)}$ of a Wilson loop in the theory with two $SU(N)$ vector multiplets. Using the definition \eqref{Wlmm} and expanding $\mathcal{S}_{\text{int}}$ we have
\begin{equation}\begin{split}\label{examplew}
    w_{[1]}^{(2)}&=\frac{1}{N} \sum_{\ell = 0}^{\infty}\frac{1}{\ell!} \parenth{\frac{{\color{blue}\lambda}_1}{2 N}}^{\frac{\ell}{2}}\langle \tr a_1^\ell\rangle_2\\
    &=\!\frac{1}{N}\! \sum_{\ell = 0}^{\infty}\frac{1}{\ell!}\! \parenth{\!\frac{{\color{blue}\lambda}_1}{2 N}\!}^{\!\!\frac{\ell}{2}}\!\!\left[t^{(1)}_{[\ell]}
 \!+\!\frac{3{\color{blue}\lambda}_1{\color{red}\zeta}_3}{64\pi^4N^2}\!\left[(t^{(1)}_{[\ell]}t^{(1)}_{[2,2]}-t^{(1)}_{[\ell,2,2]}){\color{blue}\lambda}_1\!-\!2{\color{blue}\lambda}_2(t^{(1)}_{[\ell]}t^{(1)}_{[2]}-t^{(1)}_{[\ell,2]})t^{(2)}_{[2]}\right]\!\!+\!...\right]
\end{split}\end{equation}
where we used the definition \eqref{rectn}.

\underline{Option a}: Let's consider for instance the perturbative expansion of \eqref{examplew} up to order $\lambda^3$. We have to evaluate the first 7 elements of the sum in $\ell$ obtaining a combination of simple t-functions at any order in $\lambda$. Computing the t-functions using the recursion relation \eqref{recursionSUN}, we obtain
\begin{equation}\small\begin{split}\label{w1test1}
    w_{[1]}^{(2)}=&\;1+\frac{\left(N^2-1\right){\color{blue}\lambda}_1}{8 N^2}+\frac{\left(2 N^4-5 N^2+3\right) {\color{blue}\lambda}_1^2}{384 N^4}\\
    &+\frac{\left(N^2-1\right) {\color{blue}\lambda}_1^2 \left(\pi ^4 \left(N^4-3 N^2+3\right) {\color{blue}\lambda}_1-54 N^2 {\color{red}\zeta}_3 \left(\left(N^2+1\right) {\color{blue}\lambda}_1-(N^2-1){\color{blue}\lambda}_2\right)\right)}{9216 \pi ^4 N^6}+...\,,
\end{split}\end{equation}
where dots stand for higher orders in the couplings.

\underline{Option b}: In this case, interpreting the expansion \eqref{examplew} as a sum of terms at some fixed transcendentality, we reverse the logic used in Option a. Indeed, first we use the recursion relation \eqref{recursionSUN} to reduce the t-functions dependence on $\ell$ only as follows
\begin{align}\label{texamplesqcd}
   t_{[\ell,2,2]}&\!\!=\tfrac{1}{4}((\ell+N^2)^2-1)t_{[\ell]},\\
   t_{[\ell,2,4]}&\!\!=\!\tfrac{(\ell\!+\!N^2\!+\!3)}{16 N^2}\!\Big[\ell(\ell\!-\!1)
   (\ell\!+\!N^2\!\!+\!3)t_{[\ell-2]}\!+\!2N(\!(2\ell\!-\!5)N^2\!\!-\!(\ell\!-\!1)(\ell\!+\!3)\!+\!2N^4)t_{[\ell]}\!+\!2\ell N t_{[\ell+2]}\!\Big].\nonumber
\end{align}
Then, where it is needed as the second line of \eqref{texamplesqcd}, we shift $\ell$ in order to end up with $t_{[\ell]}$ only. Now, using \eqref{derdef}, one can rewrite the sum in $\ell$ as a combination of derivatives of the $\mathcal{N}=4$ SYM Wilson loop. For instance, the term proportional to $t^{(1)}_{[\ell,2,2]}$ in \eqref{examplew} becomes
\begin{equation}\label{tlexample}
\frac{1}{N} \sum_{\ell = 0}^{\infty}\frac{1}{\ell!} \parenth{\frac{{\color{blue}\lambda}_1}{2 N}}^{\frac{\ell}{2}}t^{(1)}_{[\ell,2,2]}=
{\color{blue}\lambda}_1^2\partial_{1}^2w_1+(N^2+1){\color{blue}\lambda}_1\partial_{1}w_1+\frac{N^4-1}{4}w_1~,
\end{equation}
where $w_1$ is given by the first line of \eqref{wIexactN=4} and the derivatives act directly on it since we used the chain rule. Repeating the same procedure for the other t-function and plugging everything in \eqref{examplew} we obtain
\begin{equation}\label{w1test2}
    w_{[1]}^{(2)}=
    w_1
    -\frac{3{\color{red}\zeta}_3{\color{blue}\lambda}_1^2}{64\pi^4N^2}
    \bigg[{\color{blue}\lambda}_1^2\partial_1^2w_1+((N^2+1){\color{blue}\lambda}_1-(N^2-1){\color{blue}\lambda}_2)\partial_1w_1\bigg]+...\,,
\end{equation}
where dots stand for higher transcendentality terms.

\paragraph{Algorithm 3: Correlators of a local operator and Wilson loops $\mathcal{A}^{(\vec{I},J)}_{\vec n}$} $\\$

The computation of $\mathcal{A}^{(\vec{I},J)}_{\vec n}$ can be summarized as the combination of the two previous algorithms
\begin{equation}
    \text{Algorithm 3}=\text{Algorithm 1}+\text{Algorithm 2}\,.
\end{equation}
Let's consider, for example, $\mathcal{A}^{(1,2)}_{[4]}$ namely the correlator of a Wilson loop belonging to the first node of the quiver and the operator $:\mathcal{O}_{[4]}^{(2)}:$ in the theory $A_1$ with gauge group $SU(N)$. Given the definition of the normal-ordered operator in the second line of \eqref{O4NO}, we have
\begin{equation}\label{A124}
    \mathcal{A}^{(1,2)}_{[4]}=    \vev{\mathcal{W}_{[1]}:\mathcal{O}_{[4]}^{(2)}:}_2=\langle\mathcal{W}_{[1]}\tilde{\mathcal{O}}_{[4]}^{(2)}\rangle_2+\alpha_{[4],[2]}^{(2,1)}\langle\mathcal{W}_{[1]}\tilde{\mathcal{O}}_{[2]}^{(1)}\rangle_2+\alpha_{[4],[2]}^{(2,2)}\langle\mathcal{W}_{[1]}\tilde{\mathcal{O}}_{[2]}^{(2)}\rangle_2\,,
\end{equation}
where the Gram-Schmidt coefficients $\alpha$ are the ones computed in \eqref{alphatest} exchanging the nodes indices $1\leftrightarrow 2$. 

Correlation functions of Wilson loops and local operators on the sphere appearing in \eqref{A124} can be computed following the same logic of Algorithm 2. Depending on the choice of Option a or b for the algorithm, one can have the observables expanded in terms of the coupling or the transcendental functions. Choosing Option b, for instance we have
\begin{equation}\begin{split}\label{SphereWOtest}
\langle\mathcal{W}_{[1]}\tilde{\mathcal{O}}_{[2]}^{(1)}\rangle_2={\color{blue}\lambda}_1 \partial_1w_1-\frac{3 {\color{red}\zeta}_3 {\color{blue}\lambda}_1^2}{64 \pi ^4 N^2} \biggl[ &{\color{blue}\lambda}_1^3\partial_1^3w_1+\left(\left(N^2+5\right) {\color{blue}\lambda}_1-(N^2-1){\color{blue}\lambda}_2\right) {\color{blue}\lambda}_1\partial_1^2w_1\\
&+\left(3 \left(N^2+1\right) {\color{blue}\lambda}_1-2 \left(N^2-1\right) {\color{blue}\lambda}_2\right) \partial_1w_1\biggl]+...\,.
\end{split}\end{equation}
Computing the remaining correlators on the sphere and plugging all together, we obtain
\begin{equation}\small\begin{split}\label{A124optb}
&\mathcal{A}^{(1,2)}_{[4]}=\frac{{\color{blue}\lambda}_1^6}{262144 \pi ^8 N^{10}}\biggl[4480  N^4\left(N^6-7 N^4+24 N^2-18\right) {\color{red}\zeta}_7  {\color{blue}\lambda}_1 \partial_1^3 w_1+16\left(N^2-1\right)N^2\\
&\times\left(420 {\color{red}\zeta}_7 N^6\!\!+\!\left(72 {\color{red}\zeta}_3^2 {\color{blue}\lambda}_1\!-\!35 {\color{red}\zeta}_7 \left({\color{blue}\lambda}_1\!+\!72\right)\right)\! N^4\!\!+\!6 \left(35 {\color{red}\zeta}_7 \left({\color{blue}\lambda}_1\!+\!36\right)\!-\!18 {\color{red}\zeta}_3^2 {\color{blue}\lambda}_1\right) N^2\!-\!630 {\color{red}\zeta}_7 {\color{blue}\lambda}_1\right) \partial_1^2w_1\\
&+\!35\! \left(N^2\!-\!1\right)^2 \left(N^4\!-\!6 N^2\!+\!18\right) {\color{red}\zeta}_7  w_1\!-\!70 \left(N^6\!-\!7 N^4\!+\!24 N^2\!-\!18\right) {\color{red}\zeta}_7 \left(8 N^4\!-\!8 N^2\!-\!{\color{blue}\lambda}_1\right) \partial_1w_1\biggl]+...
\end{split}\normalsize\end{equation}
where, since the output is cumbersome, we have considered the correlator at the orbifold point where all the couplings are equal.
In \eqref{SphereWOtest} and \eqref{A124optb}, $...$ stand for higher transcendental terms.
For completeness, the same quantity as an expansion on the couplings (Option a) reads
\begin{equation}\label{A124opta}
\mathcal{A}^{(1,2)}_{[4]}=\frac{{\color{blue}\lambda}_1^7 \left(N^2-1\right)^2}{6291456 \pi ^8 N^{12}} \left(35 {\color{red}\zeta}_7 \left(N^4-6 N^2+18\right)^2+72 \left(2 N^3-3 N\right)^2 {\color{red}\zeta}_3^2\right)+...\,,
\end{equation}
where this time $...$ represents higher orders in ${\color{blue}\lambda}_1$.

\section{QUICK manual}\label{sec:QUICK}

The main purpose of the \texttt{QUICK} package is to provide tools to automatise the computation of the observables $w$, $\mathcal{G}$ and $\mathcal{A}$ defined in \eqref{WvecImm}, \eqref{Gnormord} and \eqref{Adef} by means of the matrix model techniques introduced in section \ref{sec:MM}. The package is included in the ancillary files of this manuscript and it can also be downloaded from the GitHub repository at the following address

 \url{https://github.com/miciosca/QUICK}
 
where, in case of updates, the latest version will be loaded. In order to load the package in a Mathematica session, one has to save the file \texttt{QUICK.wl} in the same directory of the notebook and run the command 

\vspace{.1cm}
\noindent\hspace{.5cm}\includegraphics[scale=.8]{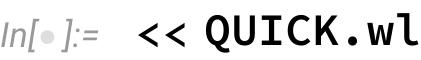}
\vspace{.1cm}

More detailed instructions, such as making the package loadable from any directory path, can be found in \cite{Preti:2017fjb,Preti:2018vog,Preti:2019rcq}. 

\subsection{The new functions}

Once \texttt{QUICK.wl} is loaded, the list of all the new functions included in the package can be shown running\footnote{Every function is equipped with a brief usage manual that can be shown running the name of the function preceded by \texttt{?}.}

\vspace{.1cm}
\noindent\hspace{.5cm}\includegraphics[scale=.8]{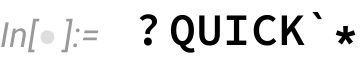}
\vspace{.1cm}

and it reads
\begin{itemize}
\item \texttt{ComputeOO[Quiver,PerturbativeOrder][\{Node1,dim1\},\{Node2,dim2\}]}: 
it computes the correlator $\mathcal{G}^{(\texttt{Node1},\texttt{Node2})}_{\texttt{dim1}}
({\color{blue}\lambda}_1,...,{\color{blue}\lambda}_\texttt{Quiver},N)$ defined in \eqref{Gnormord} up to order $\lambda^{\texttt{PerturbativeOrder}/2}$. In this case $\texttt{dim2}=\texttt{dim1}$ since normal-ordered operators are orthogonal.
\item \texttt{ComputeW[Quiver,PerturbativeOrder][NodesWL]}: it computes the expectation value of Wilson loops  $w^{(\texttt{Quiver})}_{\texttt{NodesWL}}
~({\color{blue}\lambda}_1,...,{\color{blue}\lambda}_\texttt{Quiver}~,N)$ defined in \eqref{WvecImm} up to order $\lambda^{\texttt{PerturbativeOrder}/2}$.
\item \texttt{ComputeWO[Quiver,PerturbativeOrder][NodesWL,\{Node,dim\}]}: 
it computes the correlator $\mathcal{A}^{(\texttt{NodesWL},\texttt{Node})}_{\texttt{dim}}
({\color{blue}\lambda}_1,...,{\color{blue}\lambda}_\texttt{Quiver},N)$ defined in \eqref{Adef} up to order $\lambda^{\texttt{PerturbativeOrder}/2}$.
\item \texttt{NormalOrderedOP[Quiver][Node,dim]}: it gives the normal ordered operator $:\mathcal{O}^{(\texttt{Node})}_{\texttt{dim}}:$ in terms of the basis of operators $\tilde{\mathcal{O}}$ by means of \eqref{basis} in the theory $A_{\texttt{Quiver}-1}$.
\item \texttt{SphereCorrelatorOO[Quiver,PerturbativeOrder][\{Node1,dim1\},\{Node2,dim2\}]}: 
it computes the multi matrix model correlator $\langle\tilde{\mathcal{O}}^{(\texttt{Node1})}_{\texttt{dim1}}\tilde{\mathcal{O}}^{(\texttt{Node2})}_{\texttt{dim2}}\rangle_{\texttt{Quiver}}$ up to $\lambda^{\texttt{PerturbativeOrder}/2}$.
\item \texttt{SphereCorrelatorWO[Quiver,PerturbativeOrder][NodesWL,\{Node,dim\}]}:
it computes the multi matrix model correlator $\langle\mathcal{W}_{\texttt{NodesWL}}\tilde{\mathcal{O}}^{(\texttt{Node})}_{\texttt{dim}}\rangle_{\texttt{Quiver}}$ up to $\lambda^{\texttt{PerturbativeOrder}/2}$.
\item \texttt{GramSchmidtCoeff[Quiver,PerturbativeOrder][Node,dim]}:
it computes the Gram-Schmidt coefficients appearing in the expansion of the operator $:\mathcal{O}_{\texttt{dim}}^{(\texttt{Node})}:$ by means of \eqref{basis} and \eqref{coeffGS} in the theory $A_{\texttt{Quiver}-1}$ up to $\lambda^{\texttt{PerturbativeOrder}/2}$.
\item \texttt{QUICKsaveData["filename.mx"]}:
it generates a file \texttt{filename.mx} containing the list of the internal functions already computed in the current Mathematica session as t-functions \eqref{rectn}, matrix model correlators on the sphere and Gram-Schmidt coefficients. This file can be loaded and then updated at any new session to speed-up the computational time.
\item \texttt{QUICKsaveResults["filename.mx"]}: it generates a file \texttt{filename.mx} containing the list of the perturbative expansions of $w$, $\mathcal{G}$ and $\mathcal{A}$ computed in the current Mathematica session. This file can be loaded and then updated at any new session to speed-up the computational time.
\end{itemize}

The functions listed above depends on few inputs with the following syntax.   
\texttt{Quiver} is a non-negative integer that specifies the number of nodes $q$ of the considered quiver theory $A_{q-1}$. \texttt{PerturbativeOrder} is the maximum order in the perturbative expansion of the observable. It is a non-negative integer and, depending on the case, it corresponds either to the maximal power of $g_I=\sqrt{\lambda}/N$ or the order of expansion of the matrix model action rearranged as a transcendental expansion. It ignores the normalisation of the operators \eqref{Ondefinition} even if this is included in the final results.
For a selected local operator, \texttt{Node} and \texttt{dim} stand for the label $I$ of the vector multiplet in which it belongs and its dimension $\vec{n}$ respectively. It is understood that when they appear followed by a number, they refer to different local operators. Notice that, since operators \eqref{Ondefinition} in general are multi-trace, \texttt{dim} is a \texttt{List} of all the powers $\{n_1,n_2,...\}$ appearing in the operator, while \texttt{Node} is an integer number in the interval $[1,\texttt{Quiver}]$. In presence of Wilson loops, \texttt{NodesWL} represents the vector $\vec{I}$ of nodes of the quiver in which the Wilson loops belong. Its syntax is a \texttt{List} of labels $\{I_1,I_2,...\}$ that reduces to a single element $\{I\}$, in case of only one Wilson loop considered.
Finally, \texttt{"filename.mx"} is a \texttt{String}.

Besides the functions listed above, the package offers four additional options. Those options admits two possible values: \texttt{True} or \texttt{False} and they can be modified at any time in the Mathematica session. Their purpose is to provide more freedom of choice of the main parameters of the theory and they reads 
\begin{itemize}
    \item \texttt{\$UNgroup}: it specifies the gauge group of the considered theory. It is \texttt{False} for $SU(N)$ gauge group and \texttt{True} for $U(N)$ gauge group. It is initialised as \texttt{False}.
    \item \texttt{\$LargeN}: it specifies if the theory is considered in the planar limit or not. If \texttt{False}, the rank $N$ of the gauge group is kept finite, if \texttt{True}, the limit $N\rightarrow \infty$ is turned on\footnote{Not all the functions are affected by this option. Indeed, in some intermediate steps, $N$ needs to be finite and then set to infinity at the very end of the computation of correlators).}. It is initialised as \texttt{False}.    
    \item \texttt{\$OrbifoldPoint}: it specifies if the theory is considered at the orbifold point or not. It is \texttt{False} if all the coupling constants $\lambda_1,...,\lambda_q$ are kept independent, while it is \texttt{True} at the orbifold point. It is initialised as \texttt{False}.
    \item \texttt{\$TranscendentalExp}: In case of correlators that include Wilson loops, it switches between Option a and Option b of Algorithm 2 and Algorithm 3 (see section \ref{sec:algorithm}). If it is \texttt{False}, correlators are expanded in the coupling constants (Option a), if it is \texttt{True}, correlators are expanded in transcendental functions (Option b). It is initialised as \texttt{True}.
\end{itemize}

Any of the function listed above strongly depends on the computation of auxiliary objects such as the Gram-Schmidt coefficients, two-point functions on the $S^4$ sphere but above all the $t$-functions \eqref{rectn}. Indeed, the recursion relations \eqref{recursionSUN} and \eqref{recursionUN} generate several $t$-functions depending on the length of the starting ones. In order to increase the package efficiency, all the pre-computed $t$-functions as well as the Gram-Schmidt coefficients and the sphere correlators are stored in memory and their value is automatically used in case they will appear in subsequent computations. Similarly, if some of the pre-computed data are needed at a higher order in perturbation theory, to improve the computation time only the missing orders are computed and then updated in memory.
In case in which it is necessary to close the Mathematica session, it is convenient to export the whole set of pre-compute data using the function \texttt{QUICKsaveData} and then re-load the database in a new session. Finally, the code is highly parallelised, then the efficiency of the package is directly proportional to the number of available parallel kernels in the Mathematica session. 

\subsection{Examples}

In this section we present some examples of usage of the package reproducing the results appearing in the previous sections. This is a very limited set of examples that show only partially the possible use of the package. A more detailed set of applications is presented in the ancillary notebook  "\texttt{QUICKExample.nb}" attached to this manuscript. 

Let's start computing the correlator $\mathcal{G}_{[4]}^{(1,2)}$ with the Algorithm 1 presented in section \ref{sec:algorithm}. The considered operator is a single-trace of length 4 defined by \eqref{Onmmdefinition} with normalisation $C^{(1)}_{[4]}=\frac{\lambda_1^2}{(8\pi^2N)^2}$. Its normal ordering, in case $U(N)$ is the chosen gauge group, is given by the following command

\vspace{.1cm}
\noindent\hspace{.1cm}\includegraphics[scale=.65]{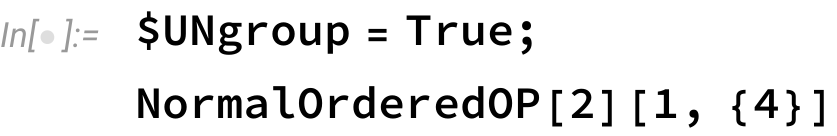}\\
\noindent\includegraphics[scale=.63]{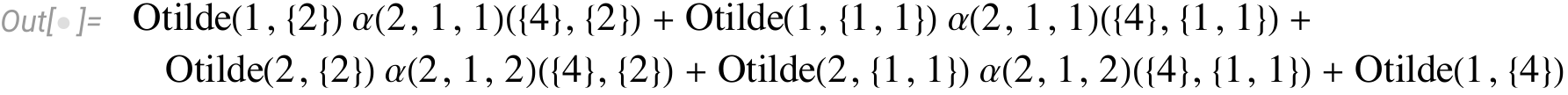}
\vspace{.1cm}

while in case of $SU(N)$ gauge group, the \texttt{\$UNgroup} option has to be changed accordingly, then the normal ordered operator is given by 

\vspace{.1cm}
\noindent\hspace{.1cm}\includegraphics[scale=.65]{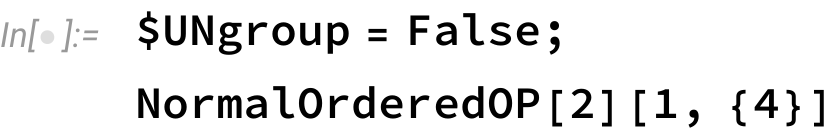}\\
\noindent\includegraphics[scale=.63]{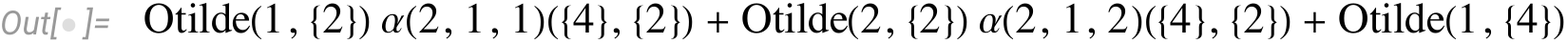}
\vspace{.1cm}

in agreement with \eqref{O4NO} (the notation is self-explanatory). Since in section \ref{sec:algorithm} we chose $SU(N)$ as gauge group, for the rest of this section we will keep \texttt{\$UNgroup=False}. The next step of the algorithm is to express the correlator $\mathcal{G}_{[4]}^{(1,2)}$ in terms of the Gram-Schmidt coefficients and the matrix model two-point functions as in \eqref{G4}. The latter can be easily computed using the function \texttt{SphereCorrelatorOO}, for instance we have

\vspace{.1cm}
\noindent\hspace{.1cm}\includegraphics[scale=.6]{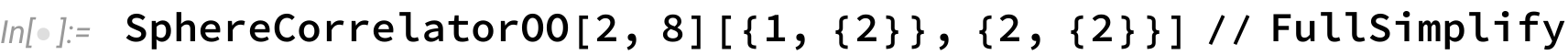}\\
\noindent\includegraphics[scale=.58]{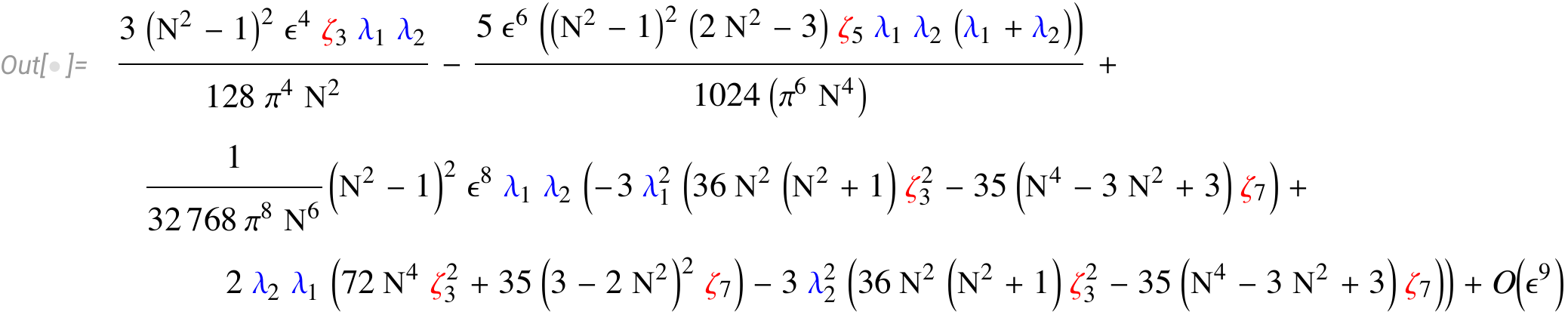}
\vspace{.1cm}

as shown in \eqref{sphereOOtest}.
Notice that, in order to simplify the expansion in presence of several coupling constants, 
we re-scale all the couplings as $\lambda_I\rightarrow \epsilon^2 \lambda_I$ and then we expand for $\epsilon\rightarrow 0$\footnote{All the outputs of the functions of \texttt{QUICK} that are written as perturbative expansions are written with this re-scaled couplings and expanded in $\epsilon$.}.
Finally, it is also possible to compute the Gram-Schmidt coefficients using the following command 

\vspace{.1cm}
\noindent\hspace{.1cm}\includegraphics[scale=.6]{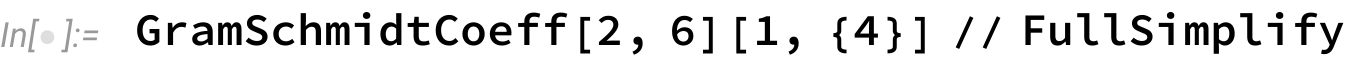}\\
\noindent\includegraphics[scale=.58]{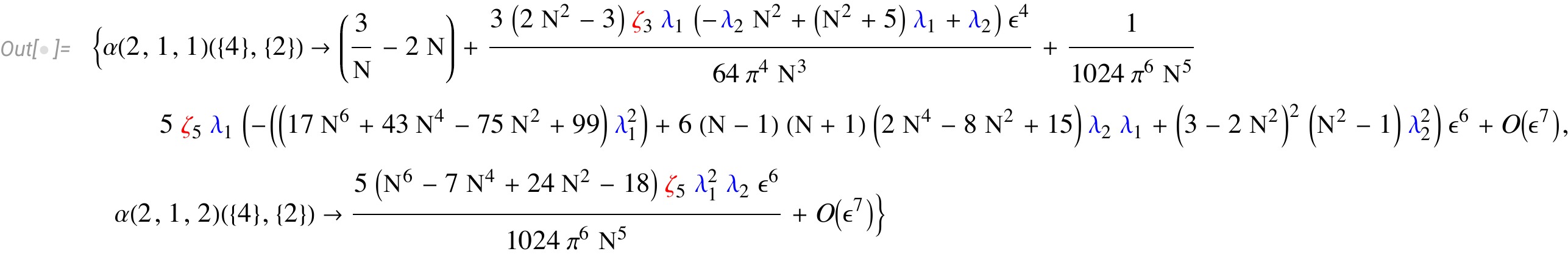}
\vspace{.1cm}

that reproduces exactly \eqref{alphatest} once the two-point functions on the sphere are substituted by their expansions in terms of the couplings.

All the previous steps are implemented and optimised in the function \texttt{ComputeOO}. Then, the correlator $\mathcal{G}_{[4]}^{(1,2)}$ can be simply computed using the following syntax

\vspace{.1cm}
\noindent\hspace{.1cm}\includegraphics[scale=.6]{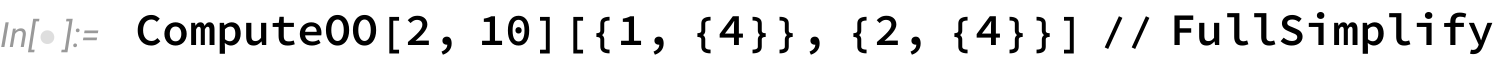}\\
\noindent\includegraphics[scale=.58]{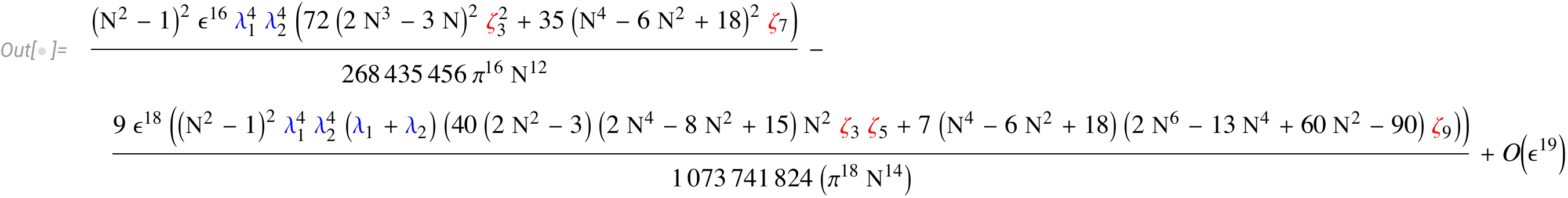}
\vspace{.1cm}

in agreement with \eqref{G124}.

Let's consider now the expectation value of the Wilson loop $w_{[1]}^{(2)}$ as presented in the Algorithm 2 in section \ref{sec:algorithm}. In order to compute it as an expansion in the couplings (Option a) and compare it with \eqref{w1test1}, one has to use the following syntax

\vspace{.1cm}
\noindent\hspace{.1cm}\includegraphics[scale=.6]{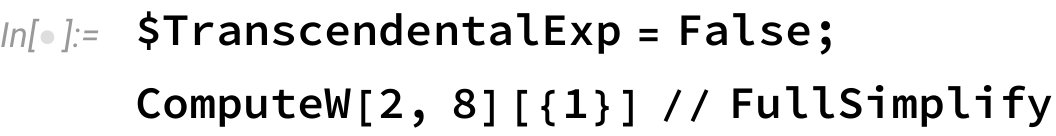}\\
\noindent\includegraphics[scale=.525]{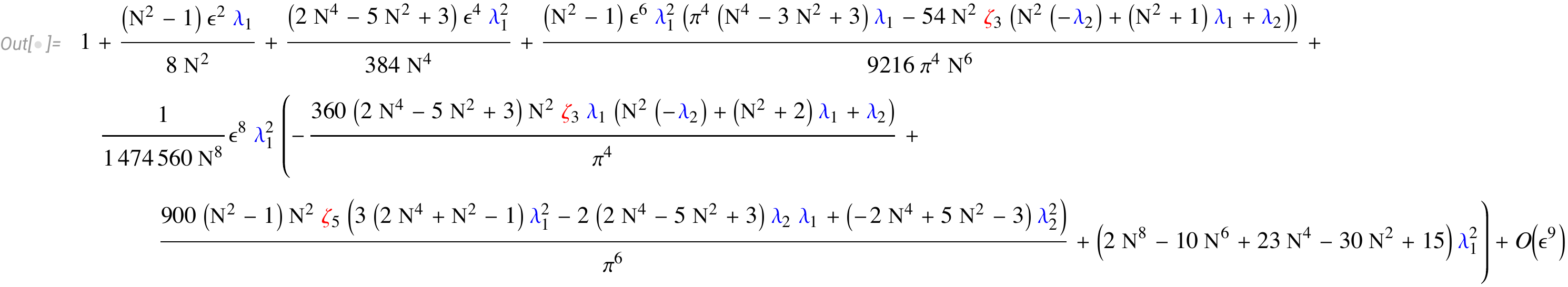}
\vspace{.1cm}

while to compute it in terms of an expansion in transcendental functions (Option b), one has to change the value of \texttt{\$TranscendentalExp} to \texttt{True} and re-run the command

\vspace{.1cm}
\noindent\hspace{.1cm}\includegraphics[scale=.6]{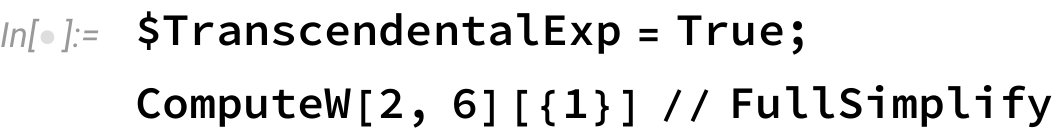}\\
\noindent\includegraphics[scale=.58]{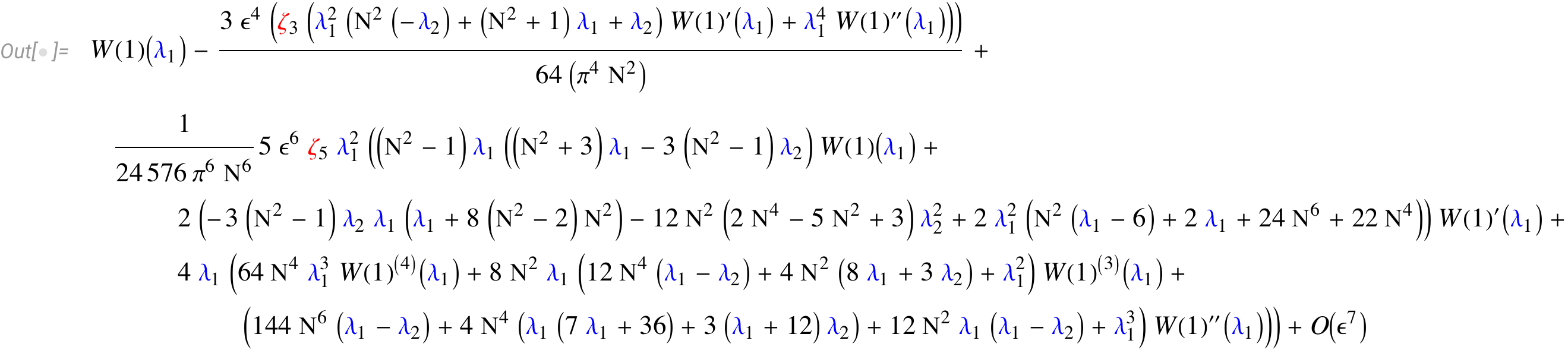}
\vspace{.1cm}

in agreement with \eqref{w1test2}. In the output we use the following notation for the derivatives of the $\mathcal{N}=4$ SYM Wilson loops 
\begin{equation}
\mathtt{W(n)^{(k)}({\color{blue}\lambda}_I)}=\frac{d^k}{d{\color{blue}\lambda}_I^k}w_{[\underbrace{I,...,I}_{n-times}]}\,,
\end{equation}
where $w_I$ is given by \eqref{wIexactN=4}, $w_{[I,I]}$ by \eqref{wIIexactN=4} and so on. 

However, there are some cases in which it is not possible to re-cast all the terms of the transcendental expansion into derivatives of $\mathcal{N}=4$ SYM Wilson loops. This is the case, for instance, of expectation values containing coincident Wilson loops belonging to the same vector multiplet. A simple example is the following observable

\vspace{.1cm}
\noindent\hspace{.1cm}\includegraphics[scale=.6]{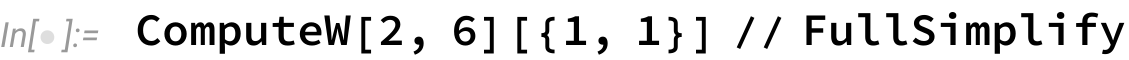}\\
\noindent\includegraphics[scale=.58]{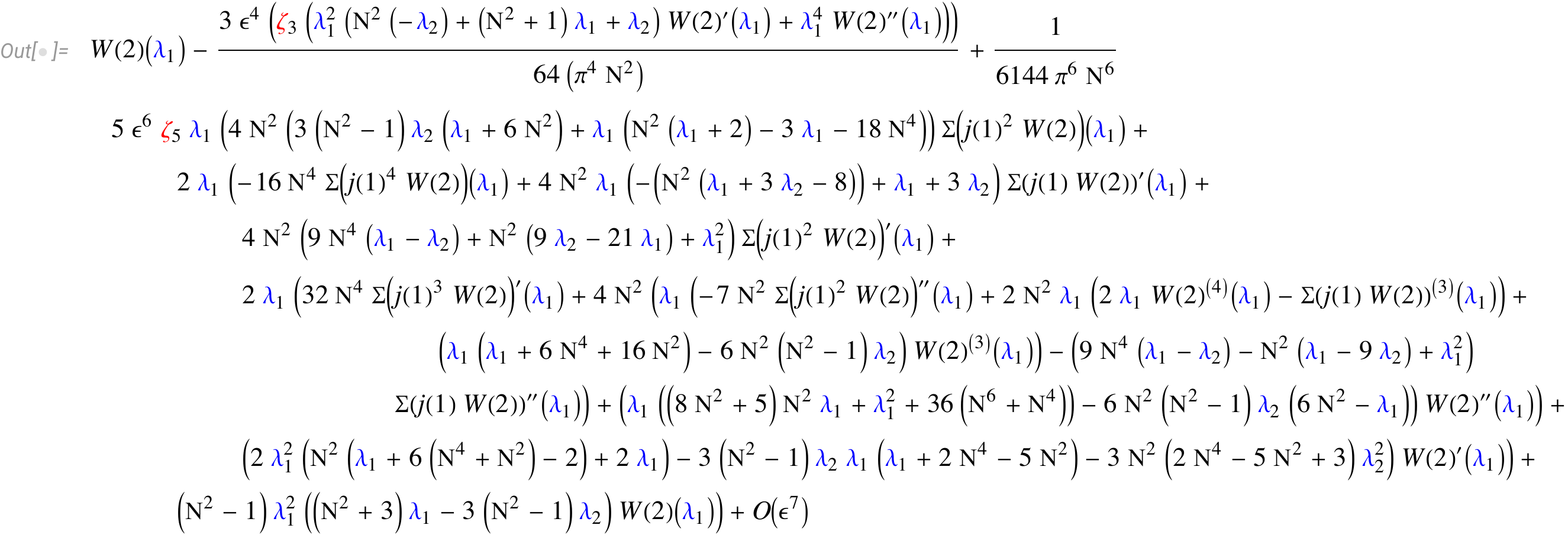}
\vspace{.1cm}

Those quantities can be expressed in terms of derivatives of certain infinite sums and they appears in the output with the following notation
\begin{equation}\small\begin{split}
\mathtt{\Sigma(j(1)^{m_1}...j(n-1)^{m_{n-1}}W(n))^{(k)}({\color{blue}\lambda}_I)}=
\frac{1}{N^n} \frac{d^k}{d{\color{blue}\lambda}_I^k}\sum_{\ell_1 ,\ell_2,...,\ell_n }\frac{\ell_1^{m_1}...\ell_{n-1}^{m_{n-1}}}{\ell_1!...\ell_n!} \parenth{\frac{{\color{blue}\lambda}_{I}}{2 N}}^{\frac{\ell_1+...+\ell_n}{2}}t_{[\ell_1,\ell_2,...,\ell_n]}.
\end{split}\normalsize\end{equation}
Considering how involved are the sums above, in these cases it is convenient to turn off the option \texttt{\$TranscendentalExp} and compute the vev in terms of a series in the couplings.

The last examples concerns the correlator $\mathcal{A}_{[4]}^{(1,2)}$ 
computed with the Algorithm 3. We again consider a single-trace operator of length 4 defined by \eqref{Onmmdefinition} with normalisation $C^{(1)}_{[4]}=\lambda_1/N$. The local operator belongs to the vector multiplet labelled by $I=2$ while the Wilson loop belongs to the first node. Their correlation function can be written as \eqref{A124}. The Gram-Schmidt coefficients are already pre-computed in the previous examples, while the two-point functions on the sphere can be computed by the following command

\vspace{.1cm}
\noindent\hspace{.1cm}\includegraphics[scale=.6]{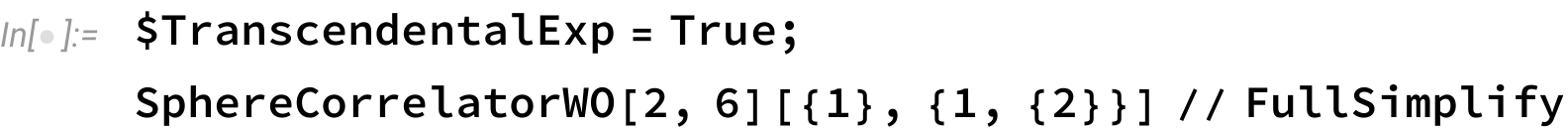}\\
\noindent\includegraphics[scale=.58]{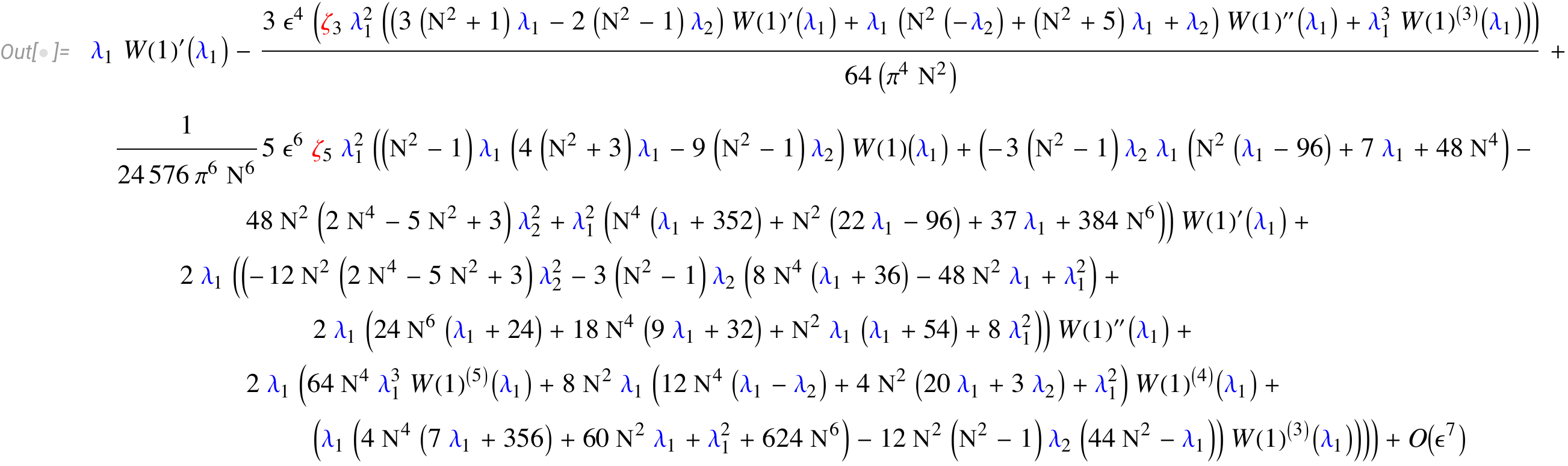}
\vspace{.1cm}

in agreement with \eqref{SphereWOtest}. The package provides the function \texttt{ComputeWO} that directly computes the correlators $\mathcal{A}$ without any additional input. Then, the observables we are considering is given by 

\vspace{.1cm}
\noindent\hspace{.1cm}\includegraphics[scale=.6]{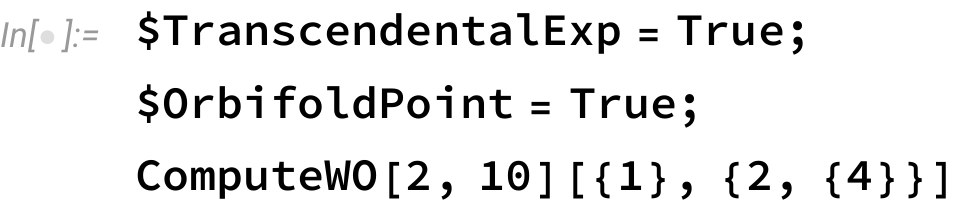}\\
\noindent\includegraphics[scale=.58]{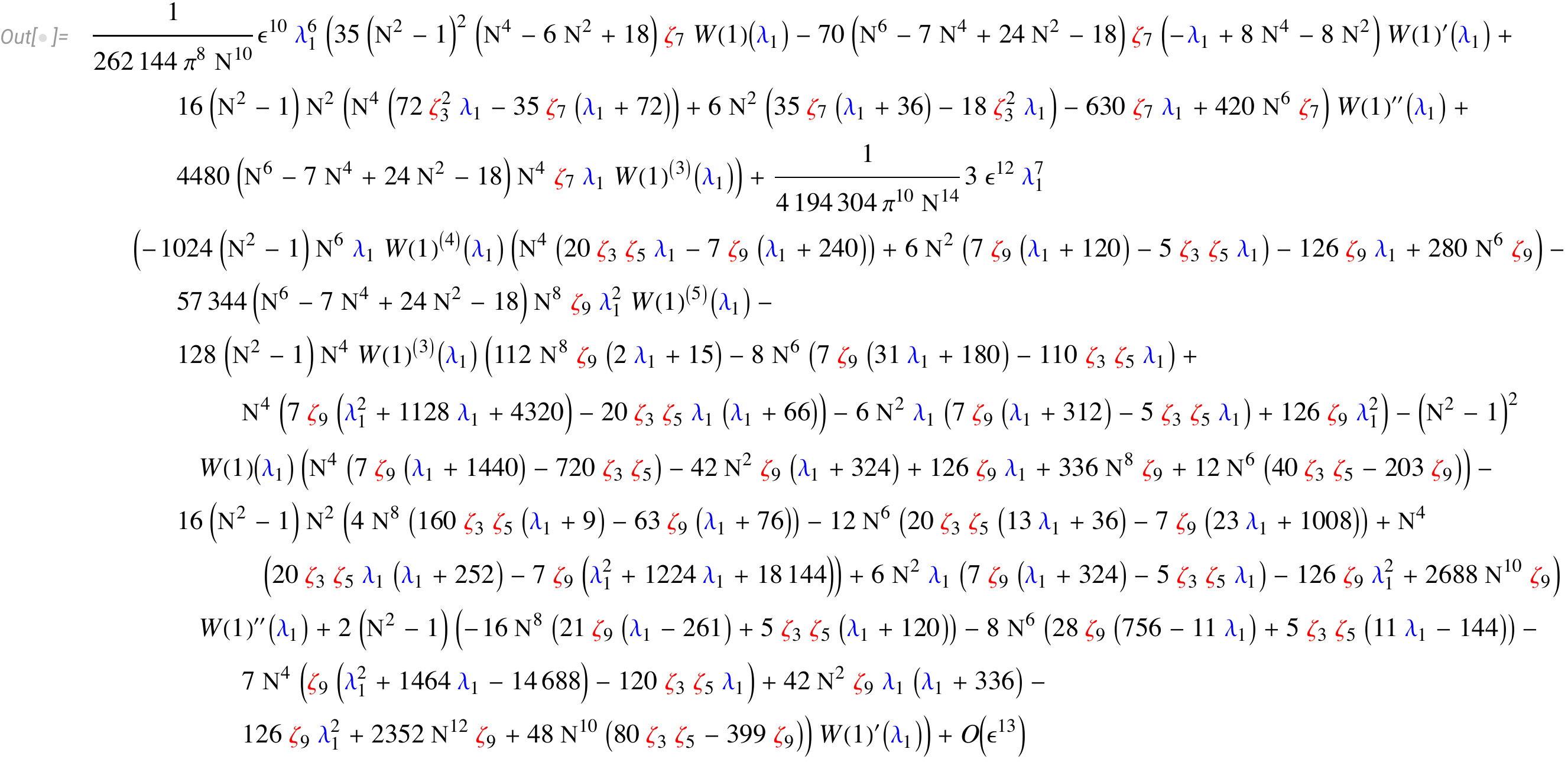}
\vspace{.1cm}

as an expansion in transcendental functions (Option b) and it is given by

\vspace{.1cm}
\noindent\hspace{.1cm}\includegraphics[scale=.6]{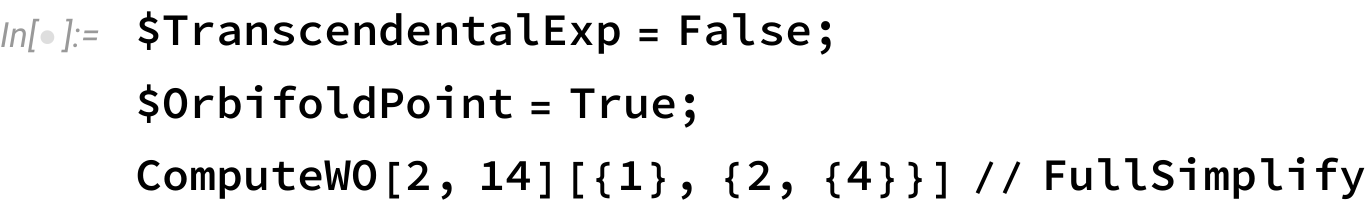}\\
\noindent\includegraphics[scale=.58]{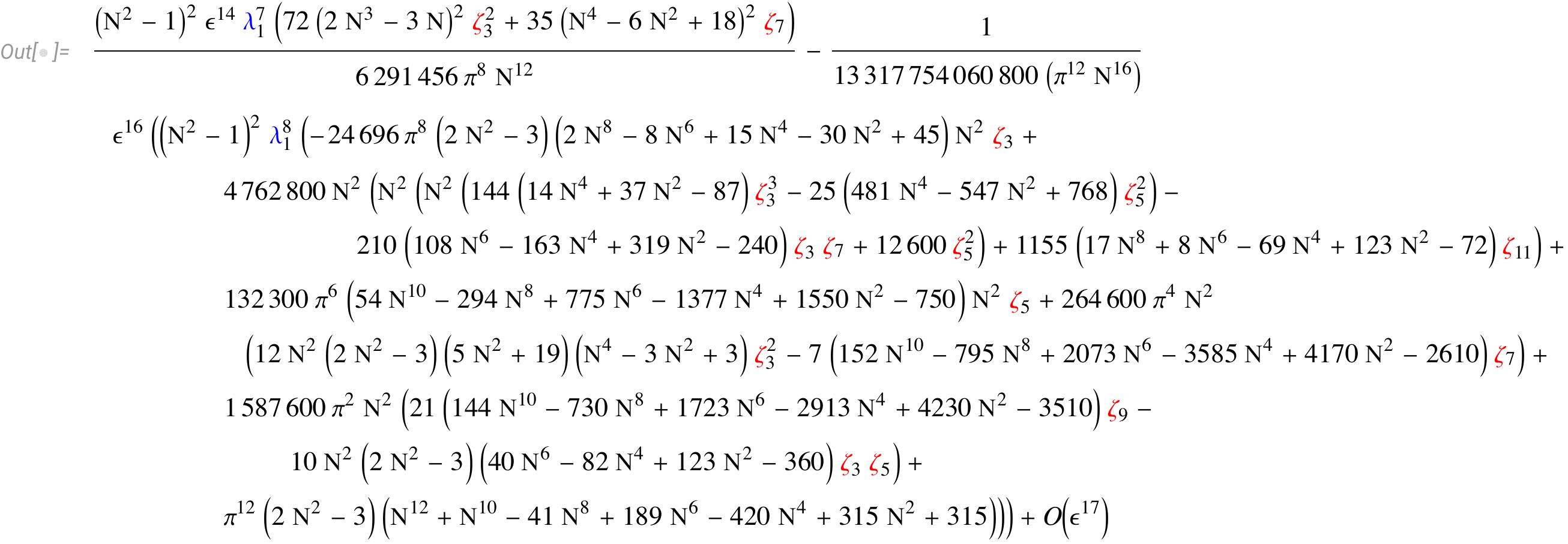}
\vspace{.1cm}

as an expansion in the coupling constants. In both cases, we considered the theory at the orbifold point to compare with 
\eqref{A124optb} and \eqref{A124opta}.

We want to stress that all the algorithms of section \ref{sec:algorithm} and their dedicated functions are very efficient to compute also correlators in SCQCD, $\mathcal{N}=4$ SYM and its $\mathbb{Z}_q$ orbifolds. In particular, in order to obtain observables in $\mathcal{N}=4$ SYM, we have to decouple the $\mathcal{S}_{\text{int}}$ action from \eqref{vevf} and this is equivalent to set all the Riemann zetas to zero ${\color{red}\zeta}_{i}\rightarrow 0$. Consider, for instance, the simplest $SU(N)$ correlator between a Wilson loop and a local operator belonging to the same vector multiplet. This is equivalent to the following expression

\vspace{.1cm}
\noindent\hspace{.1cm}\includegraphics[scale=.6]{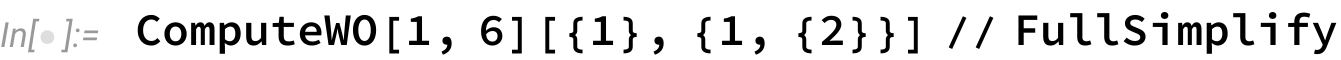}\\
\noindent\includegraphics[scale=.58]{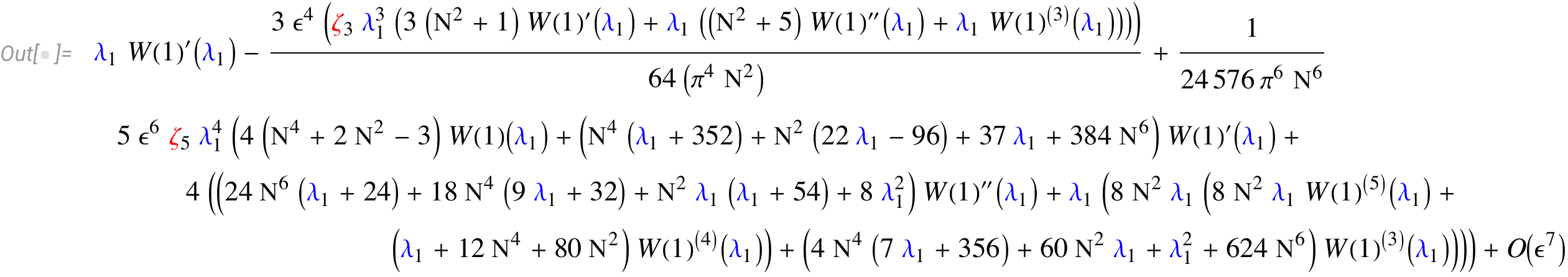}
\vspace{.1cm}

The $\mathcal{N}=4$ SYM equivalent of this observable is given by the only terms that doesn't contain transcendental functions, namely the leading order. In the present case, it is straightforward to test this statement. Indeed, since in $\mathcal{N}=4$ SYM the operator $O_{[2]}$ is the highest weight of the stress-tensor multiplet, we expect that its expectation value in presence of the 1/2 BPS Wilson loop (1d defect) to be proportional to the Bremsstrahlung function\footnote{The same result was later reproduced and further generalised from integrability in  \cite{Gromov:2012eu,Gromov:2013qga} and checked at strong and weak coupling in \cite{Sizov:2013joa,Bonini:2015fng}.} \cite{Correa:2012at,Fiol:2012sg}. In our notation\footnote{The constant of proportionality between the $\mathcal{N}=4$ SYM correlator and the $\mathcal{N}=2$ one depend on the choice of the normalisation in \eqref{Ondefinition} and \eqref{Onmmdefinition}.} we have
\begin{equation}
\vev{ W_{[1]} \, O_{[2]}^{(1)}}_{\mathcal{N}=4}=\frac{1}{4\pi^2}\mathcal{A}_{[2]}^{(1,1)}\biggl|_{{\color{red}\zeta}_{i}\rightarrow 0}
= \frac12 B(g) w_{1}\,,
\end{equation}
with the Bremsstrahlung function defined as $B(g)=\frac{1}{2\pi^2}\lambda\frac{w_{1}'}{w_{1}}$ in agreement with our result for $\mathcal{A}_{[2]}^{(1,1)}$.

\section*{Acknowledgments}
We thank M. Bill\'o and F. Galvagno for useful discussions and suggestions. 
The work of MP is supported by European Research Council (ERC) under the European Union’s Horizon
2020 research and innovation programme (grant agreement No. 865075) EXACTC.

\appendix
\section{The action of $\mathcal{N}=2$ theories}\label{sec:appendixA}
The action of $A_{q-1}$ theories can be written in the $\cN=1$ superspace formalism.
The $\cN=2$ vector field in the node $I$ is decomposed into a $\cN=1$ vector multiplet $V_I$ and a $\cN=1$ chiral multiplet $\Phi_I$. The $\cN=2$ matter hypermultiplet is given by two $\cN=1$ chiral multiplets $\big(Q, \widetilde{Q} \big)$.
For a generic $A_{q-1}$ theory, the action can be written as \eqref{action} with the gauge term given by
\begin{equation}\label{Svector}
S_{\text{vector}} = \sum_{I=1}^q \Bigg[ \frac{1}{8g_I^2} \left(\int d^4x\,d^2\theta\, \tr(W_I^\a W^I_\a)\!+\!\mathrm{h.c.}\!\right) +\!2\!\int d^4x\,d^4\theta\, \tr\!\left( e^{-2g_IV_I}\Phi_I^\dagger e^{2g_IV_I}\Phi_I\right) \Bigg]
\end{equation}
and the matter term given by
\begin{align}\label{Shyper}
S_{\text{hyper}} &= \sum_{I=1}^q \Bigg[\int d^4x\,d^4\theta\,\bigg( \tr \left( Q^\dagger e^{2g_IV_I} Q e^{-2g_{I+1}V_{I+1}}\right) + \tr \left(\widetilde{Q} e^{-2g_IV_I}\widetilde{Q}^\dagger e^{2g_{I+1}V_{I+1}} \right) \bigg) \notag \\
&+ \left(\ii \sqrt{2}g_I\!\int\!d^4x\,d^2\theta\,\widetilde{Q} \Phi_I Q +\mathrm{h.c.}\right)+ \left(\ii \sqrt{2}g_{I+1}\!\int\!d^4x\,d^2\theta\,\widetilde{Q} \Phi_{I+1} Q +\mathrm{h.c.}\right) \Bigg]
\end{align}
where $g_I$ the Yang-Mills couplings and $W^I_\a$ the super field strength of $V_I$ defined as follows
\begin{align}
W^I_\a=-\frac{1}{4}\bar{D}^2\left(e^{-2g_IV_I}D_\a e^{2g_IV_I}\right)~.
\end{align}

\bibliographystyle{nb}
\bibliography{biblio.bib}

\begin{thebibliography}{10}
\ifx\href\asklfhas\newcommand{\href}[2]{#2}\fi
\ifx\arxivref\asklfhas\newcommand{\arxivref}[2]{\href{http://arxiv.org/abs/#1}{#2}}\fi
\ifx\doiref\asklfhas\newcommand{\doiref}[2]{\href{http://dx.doi.org/#1}{#2}}\fi
\raggedright
\small
\parskip 0pt

\bibitem{Galvagno:2020cgq}
F.~Galvagno and M.~Preti,
\textit{``{Chiral correlators in $ \mathcal{N} $ = 2 superconformal
  quivers}''},
\textsf{\doiref{10.1007/JHEP05(2021)201}{JHEP~2105,~201~(2021)}},
\texttt{\arxivref{2012.15792}{arXiv:2012.15792}}.

\bibitem{Galvagno:2021bbj}
F.~Galvagno and M.~Preti,
\textit{``{Wilson loop correlators in $ \mathcal{N} $ = 2 superconformal
  quivers}''},
\textsf{\doiref{10.1007/JHEP11(2021)023}{JHEP~2111,~023~(2021)}},
\texttt{\arxivref{2105.00257}{arXiv:2105.00257}}.

\bibitem{Erickson:2000af}
J.~K.~Erickson, G.~W.~Semenoff and K.~Zarembo,
\textit{``{Wilson loops in N=4 supersymmetric Yang-Mills theory}''},
\textsf{\doiref{10.1016/S0550-3213(00)00300-X}{Nucl.~Phys.~B582,~155~(2000)}},
\texttt{\arxivref{hep-th/0003055}{hep-th/0003055}}.

\bibitem{Drukker:2000rr}
N.~Drukker and D.~J.~Gross,
\textit{``{An Exact prediction of N=4 SUSYM theory for string theory}''},
\textsf{\doiref{10.1063/1.1372177}{J.~Math.~Phys.~42,~2896~(2001)}},
\texttt{\arxivref{hep-th/0010274}{hep-th/0010274}}.

\bibitem{Pestun:2007rz}
V.~Pestun,
\textit{``{Localization of gauge theory on a four-sphere and supersymmetric
  Wilson loops}''},
\textsf{\doiref{10.1007/s00220-012-1485-0}{Commun.~Math.~Phys.~313,~71~(2012)}},
\texttt{\arxivref{0712.2824}{arXiv:0712.2824}}.

\bibitem{Semenoff:2001xp}
G.~W.~Semenoff and K.~Zarembo,
\textit{``{More exact predictions of SUSYM for string theory}''},
\textsf{\doiref{10.1016/S0550-3213(01)00455-2}{Nucl.~Phys.~B616,~34~(2001)}},
\texttt{\arxivref{hep-th/0106015}{hep-th/0106015}}.

\bibitem{Pestun:2002mr}
V.~Pestun and K.~Zarembo,
\textit{``{Comparing strings in AdS(5) x S**5 to planar diagrams: An
  Example}''},
\textsf{\doiref{10.1103/PhysRevD.67.086007}{Phys.~Rev.~D67,~086007~(2003)}},
\texttt{\arxivref{hep-th/0212296}{hep-th/0212296}}.

\bibitem{Drukker:2007qr}
N.~Drukker, S.~Giombi, R.~Ricci and D.~Trancanelli,
\textit{``{Supersymmetric Wilson loops on S**3}''},
\textsf{\doiref{10.1088/1126-6708/2008/05/017}{JHEP~0805,~017~(2008)}},
\texttt{\arxivref{0711.3226}{arXiv:0711.3226}}.

\bibitem{Pestun:2009nn}
V.~Pestun,
\textit{``{Localization of the four-dimensional N=4 SYM to a two-sphere and 1/8
  BPS Wilson loops}''},
\textsf{\doiref{10.1007/JHEP12(2012)067}{JHEP~1212,~067~(2012)}},
\texttt{\arxivref{0906.0638}{arXiv:0906.0638}}.

\bibitem{Giombi:2009ds}
S.~Giombi and V.~Pestun,
\textit{``{Correlators of local operators and 1/8 BPS Wilson loops on S**2 from
  2d YM and matrix models}''},
\textsf{\doiref{10.1007/JHEP10(2010)033}{JHEP~1010,~033~(2010)}},
\texttt{\arxivref{0906.1572}{arXiv:0906.1572}}.

\bibitem{Giombi:2009ek}
S.~Giombi and V.~Pestun,
\textit{``{The 1/2 BPS 't Hooft loops in N=4 SYM as instantons in 2d
  Yang-Mills}''},
\textsf{\doiref{10.1088/1751-8113/46/9/095402}{J.~Phys.~A46,~095402~(2013)}},
\texttt{\arxivref{0909.4272}{arXiv:0909.4272}}.

\bibitem{Giombi:2012ep}
S.~Giombi and V.~Pestun,
\textit{``{Correlators of Wilson Loops and Local Operators from Multi-Matrix
  Models and Strings in AdS}''},
\textsf{\doiref{10.1007/JHEP01(2013)101}{JHEP~1301,~101~(2013)}},
\texttt{\arxivref{1207.7083}{arXiv:1207.7083}}.

\bibitem{Bonini:2014vta}
M.~Bonini, L.~Griguolo and M.~Preti,
\textit{``{Correlators of chiral primaries and 1/8 BPS Wilson loops from
  perturbation theory}''},
\textsf{\doiref{10.1007/JHEP09(2014)083}{JHEP~1409,~083~(2014)}},
\texttt{\arxivref{1405.2895}{arXiv:1405.2895}}.

\bibitem{Bonini:2015fng}
M.~Bonini, L.~Griguolo, M.~Preti and D.~Seminara,
\textit{``{Bremsstrahlung function, leading L\"uscher correction at weak
  coupling and localization}''},
\textsf{\doiref{10.1007/JHEP02(2016)172}{JHEP~1602,~172~(2016)}},
\texttt{\arxivref{1511.05016}{arXiv:1511.05016}}.

\bibitem{Correa:2012at}
D.~Correa, J.~Henn, J.~Maldacena and A.~Sever,
\textit{``{An exact formula for the radiation of a moving quark in N=4 super
  Yang Mills}''},
\textsf{\doiref{10.1007/JHEP06(2012)048}{JHEP~1206,~048~(2012)}},
\texttt{\arxivref{1202.4455}{arXiv:1202.4455}}.

\bibitem{Lewkowycz:2013laa}
A.~Lewkowycz and J.~Maldacena,
\textit{``{Exact results for the entanglement entropy and the energy radiated
  by a quark}''},
\textsf{\doiref{10.1007/JHEP05(2014)025}{JHEP~1405,~025~(2014)}},
\texttt{\arxivref{1312.5682}{arXiv:1312.5682}}.

\bibitem{Okuyama:2018aij}
K.~Okuyama,
\textit{``{Connected correlator of 1/2 BPS Wilson loops in $\mathcal{N}=4$
  SYM}''},
\textsf{\doiref{10.1007/JHEP10(2018)037}{JHEP~1810,~037~(2018)}},
\texttt{\arxivref{1808.10161}{arXiv:1808.10161}}.

\bibitem{Correa:2018lyl}
D.~H.~Correa, P.~Pisani and A.~Rios~Fukelman,
\textit{``{Ladder Limit for Correlators of Wilson Loops}''},
\textsf{\doiref{10.1007/JHEP05(2018)168}{JHEP~1805,~168~(2018)}},
\texttt{\arxivref{1803.02153}{arXiv:1803.02153}}.

\bibitem{Correa:2018pfn}
D.~Correa, P.~Pisani, A.~Rios~Fukelman and K.~Zarembo,
\textit{``{Dyson equations for correlators of Wilson loops}''},
\textsf{\doiref{10.1007/JHEP12(2018)100}{JHEP~1812,~100~(2018)}},
\texttt{\arxivref{1811.03552}{arXiv:1811.03552}}.

\bibitem{CanazasGaray:2019mgq}
A.~F.~Canazas~Garay, A.~Faraggi and W.~M\"uck,
\textit{``{Note on generating functions and connected correlators of 1/2-BPS
  Wilson loops in $\mathcal{N}=4$ SYM theory}''},
\textsf{\doiref{10.1007/JHEP08(2019)149}{JHEP~1908,~149~(2019)}},
\texttt{\arxivref{1906.03816}{arXiv:1906.03816}}.

\bibitem{Beccaria:2020ykg}
M.~Beccaria and A.~A.~Tseytlin,
\textit{``{On the structure of non-planar strong coupling corrections to
  correlators of BPS Wilson loops and chiral primary operators}''},
\textsf{\doiref{10.1007/JHEP01(2021)149}{JHEP~2101,~149~(2021)}},
\texttt{\arxivref{2011.02885}{arXiv:2011.02885}}.

\bibitem{Panerai:2018ryw}
R.~Panerai, M.~Poggi and D.~Seminara,
\textit{``{Supersymmetric Wilson loops in two dimensions and duality}''},
\textsf{\doiref{10.1103/PhysRevD.100.025011}{Phys.~Rev.~D~100,~025011~(2019)}},
\texttt{\arxivref{1812.01315}{arXiv:1812.01315}}.

\bibitem{Kapustin:2009kz}
A.~Kapustin, B.~Willett and I.~Yaakov,
\textit{``{Exact Results for Wilson Loops in Superconformal Chern-Simons
  Theories with Matter}''},
\textsf{\doiref{10.1007/JHEP03(2010)089}{JHEP~1003,~089~(2010)}},
\texttt{\arxivref{0909.4559}{arXiv:0909.4559}}.

\bibitem{Marino:2009jd}
M.~Marino and P.~Putrov,
\textit{``{Exact Results in ABJM Theory from Topological Strings}''},
\textsf{\doiref{10.1007/JHEP06(2010)011}{JHEP~1006,~011~(2010)}},
\texttt{\arxivref{0912.3074}{arXiv:0912.3074}}.

\bibitem{Drukker:2010nc}
N.~Drukker, M.~Marino and P.~Putrov,
\textit{``{From weak to strong coupling in ABJM theory}''},
\textsf{\doiref{10.1007/s00220-011-1253-6}{Commun.~Math.~Phys.~306,~511~(2011)}},
\texttt{\arxivref{1007.3837}{arXiv:1007.3837}}.

\bibitem{Bianchi:2018bke}
M.~S.~Bianchi, L.~Griguolo, A.~Mauri, S.~Penati and D.~Seminara,
\textit{``{A matrix model for the latitude Wilson loop in ABJM theory}''},
\textsf{\doiref{10.1007/JHEP08(2018)060}{JHEP~1808,~060~(2018)}},
\texttt{\arxivref{1802.07742}{arXiv:1802.07742}}.

\bibitem{Griguolo:2021rke}
L.~Griguolo, L.~Guerrini and I.~Yaakov,
\textit{``{Localization and duality for ABJM latitude Wilson loops}''},
\textsf{\doiref{10.1007/JHEP08(2021)001}{JHEP~2108,~001~(2021)}},
\texttt{\arxivref{2104.04533}{arXiv:2104.04533}}.

\bibitem{Bianchi:2014laa}
M.~S.~Bianchi, L.~Griguolo, M.~Leoni, S.~Penati and D.~Seminara,
\textit{``{BPS Wilson loops and Bremsstrahlung function in ABJ(M): a two loop
  analysis}''},
\textsf{\doiref{10.1007/JHEP06(2014)123}{JHEP~1406,~123~(2014)}},
\texttt{\arxivref{1402.4128}{arXiv:1402.4128}}.

\bibitem{Bianchi:2017svd}
M.~S.~Bianchi, L.~Griguolo, A.~Mauri, S.~Penati, M.~Preti and D.~Seminara,
\textit{``{Towards the exact Bremsstrahlung function of ABJM theory}''},
\textsf{\doiref{10.1007/JHEP08(2017)022}{JHEP~1708,~022~(2017)}},
\texttt{\arxivref{1705.10780}{arXiv:1705.10780}}.

\bibitem{Bianchi:2017ozk}
L.~Bianchi, L.~Griguolo, M.~Preti and D.~Seminara,
\textit{``{Wilson lines as superconformal defects in ABJM theory: a formula for
  the emitted radiation}''},
\textsf{\doiref{10.1007/JHEP10(2017)050}{JHEP~1710,~050~(2017)}},
\texttt{\arxivref{1706.06590}{arXiv:1706.06590}}.

\bibitem{Bianchi:2018scb}
L.~Bianchi, M.~Preti and E.~Vescovi,
\textit{``{Exact Bremsstrahlung functions in ABJM theory}''},
\textsf{\doiref{10.1007/JHEP07(2018)060}{JHEP~1807,~060~(2018)}},
\texttt{\arxivref{1802.07726}{arXiv:1802.07726}}.

\bibitem{Drukker:2019bev}
N.~Drukker et~al.,
\textit{``{Roadmap on Wilson loops in 3d Chern\textendash{}Simons-matter
  theories}''},
\textsf{\doiref{10.1088/1751-8121/ab5d50}{J.~Phys.~A~53,~173001~(2020)}},
\texttt{\arxivref{1910.00588}{arXiv:1910.00588}}.

\bibitem{Andree:2010na}
R.~Andree and D.~Young,
\textit{``{Wilson Loops in N=2 Superconformal Yang-Mills Theory}''},
\textsf{\doiref{10.1007/JHEP09(2010)095}{JHEP~1009,~095~(2010)}},
\texttt{\arxivref{1007.4923}{arXiv:1007.4923}}.

\bibitem{Passerini:2011fe}
F.~Passerini and K.~Zarembo,
\textit{``{Wilson Loops in N=2 Super-Yang-Mills from Matrix Model}''},
\textsf{\doiref{10.1007/JHEP10(2011)065,
  10.1007/JHEP09(2011)102}{JHEP~1109,~102~(2011)}},
\texttt{\arxivref{1106.5763}{arXiv:1106.5763}},
[Erratum: JHEP10,065(2011)].

\bibitem{Bourgine:2011ie}
J.-E.~Bourgine,
\textit{``{A Note on the integral equation for the Wilson loop in N = 2 D=4
  superconformal Yang-Mills theory}''},
\textsf{\doiref{10.1088/1751-8113/45/12/125403}{J.~Phys.~A45,~125403~(2012)}},
\texttt{\arxivref{1111.0384}{arXiv:1111.0384}}.

\bibitem{Billo:2019fbi}
M.~Billo, F.~Galvagno and A.~Lerda,
\textit{``{BPS Wilson loops in generic conformal $ \mathcal{N} $ = 2 SU(N) SYM
  theories}''},
\textsf{\doiref{10.1007/JHEP08(2019)108}{JHEP~1908,~108~(2019)}},
\texttt{\arxivref{1906.07085}{arXiv:1906.07085}}.

\bibitem{Fiol:2020bhf}
B.~Fiol, J.~Mart\'\i{}nez-Montoya and A.~Rios~Fukelman,
\textit{``{The planar limit of $\mathcal{N}=2$ superconformal field
  theories}''},
\textsf{\doiref{10.1007/JHEP05(2020)136}{JHEP~2005,~136~(2020)}},
\texttt{\arxivref{2003.02879}{arXiv:2003.02879}}.

\bibitem{Beccaria:2021vuc}
M.~Beccaria, G.~V.~Dunne and A.~A.~Tseytlin,
\textit{``{BPS Wilson loop in $ \mathcal{N} $ = 2 superconformal SU(N)
  \textquotedblleft{}orientifold\textquotedblright{} gauge theory and
  weak-strong coupling interpolation}''},
\textsf{\doiref{10.1007/JHEP07(2021)085}{JHEP~2107,~085~(2021)}},
\texttt{\arxivref{2104.12625}{arXiv:2104.12625}}.

\bibitem{Rodriguez-Gomez:2016cem}
D.~Rodriguez-Gomez and J.~G.~Russo,
\textit{``{Operator mixing in large $N$ superconformal field theories on
  S$^{4}$ and correlators with Wilson loops}''},
\textsf{\doiref{10.1007/JHEP12(2016)120}{JHEP~1612,~120~(2016)}},
\texttt{\arxivref{1607.07878}{arXiv:1607.07878}}.

\bibitem{Billo:2018oog}
M.~Billo, F.~Galvagno, P.~Gregori and A.~Lerda,
\textit{``{Correlators between Wilson loop and chiral operators in $
  \mathcal{N}=2 $ conformal gauge theories}''},
\textsf{\doiref{10.1007/JHEP03(2018)193}{JHEP~1803,~193~(2018)}},
\texttt{\arxivref{1802.09813}{arXiv:1802.09813}}.

\bibitem{Beccaria:2020hgy}
M.~Beccaria, M.~Bill\`o, F.~Galvagno, A.~Hasan and A.~Lerda,
\textit{``{$ \mathcal{N} $ = 2 Conformal SYM theories at large $ \mathcal{N}
  $}''},
\textsf{\doiref{10.1007/JHEP09(2020)116}{JHEP~2009,~116~(2020)}},
\texttt{\arxivref{2007.02840}{arXiv:2007.02840}}.

\bibitem{Fiol:2015mrp}
B.~Fiol, B.~Garolera and G.~Torrents,
\textit{``{Probing $ \mathcal{N}=2 $ superconformal field theories with
  localization}''},
\textsf{\doiref{10.1007/JHEP01(2016)168}{JHEP~1601,~168~(2016)}},
\texttt{\arxivref{1511.00616}{arXiv:1511.00616}}.

\bibitem{Bianchi:2018zpb}
L.~Bianchi, M.~Lemos and M.~Meineri,
\textit{``{Line Defects and Radiation in $\mathcal{N}=2$ Conformal
  Theories}''},
\textsf{\doiref{10.1103/PhysRevLett.121.141601}{Phys.~Rev.~Lett.~121,~141601~(2018)}},
\texttt{\arxivref{1805.04111}{arXiv:1805.04111}}.

\bibitem{Bianchi:2019dlw}
L.~Bianchi, M.~Billo, F.~Galvagno and A.~Lerda,
\textit{``{Emitted Radiation and Geometry}''},
\textsf{\doiref{10.1007/JHEP01(2020)075}{JHEP~2001,~075~(2020)}},
\texttt{\arxivref{1910.06332}{arXiv:1910.06332}}.

\bibitem{Galvagno:2021qyq}
F.~Galvagno,
\textit{``{Emitted radiation in superconformal field theories}''},
\textsf{\doiref{10.1140/epjp/s13360-022-02341-2}{Eur.~Phys.~J.~Plus~137,~143~(2022)}},
\texttt{\arxivref{2112.03841}{arXiv:2112.03841}}.

\bibitem{Beccaria:2021hvt}
M.~Beccaria, M.~Bill\`o, M.~Frau, A.~Lerda and A.~Pini,
\textit{``{Exact results in a $ \mathcal{N} $ = 2 superconformal gauge theory
  at strong coupling}''},
\textsf{\doiref{10.1007/JHEP07(2021)185}{JHEP~2107,~185~(2021)}},
\texttt{\arxivref{2105.15113}{arXiv:2105.15113}}.

\bibitem{Billo:2021rdb}
M.~Billo, M.~Frau, F.~Galvagno, A.~Lerda and A.~Pini,
\textit{``{Strong-coupling results for $ \mathcal{N} $ = 2 superconformal
  quivers and holography}''},
\textsf{\doiref{10.1007/JHEP10(2021)161}{JHEP~2110,~161~(2021)}},
\texttt{\arxivref{2109.00559}{arXiv:2109.00559}}.

\bibitem{Billo:2022xas}
M.~Billo, M.~Frau, A.~Lerda, A.~Pini and P.~Vallarino,
\textit{``{Three-point functions in a $ \mathcal{N} $ = 2 superconformal gauge
  theory and their strong-coupling limit}''},
\textsf{\doiref{10.1007/JHEP08(2022)199}{JHEP~2208,~199~(2022)}},
\texttt{\arxivref{2202.06990}{arXiv:2202.06990}}.

\bibitem{Beccaria:2022ypy}
M.~Beccaria, G.~P.~Korchemsky and A.~A.~Tseytlin,
\textit{``{Strong coupling expansion in $\mathcal{N} = 2$ superconformal
  theories and the Bessel kernel}''},
\textsf{\doiref{10.1007/JHEP09(2022)226}{JHEP~2209,~226~(2022)}},
\texttt{\arxivref{2207.11475}{arXiv:2207.11475}}.

\bibitem{Kachru:1998ys}
S.~Kachru and E.~Silverstein,
\textit{``{4-D conformal theories and strings on orbifolds}''},
\textsf{\doiref{10.1103/PhysRevLett.80.4855}{Phys.~Rev.~Lett.~80,~4855~(1998)}},
\texttt{\arxivref{hep-th/9802183}{hep-th/9802183}}.

\bibitem{Gukov:1998kk}
S.~Gukov,
\textit{``{Comments on N=2 AdS orbifolds}''},
\textsf{\doiref{10.1016/S0370-2693(98)01005-3}{Phys.~Lett.~B~439,~23~(1998)}},
\texttt{\arxivref{hep-th/9806180}{hep-th/9806180}}.

\bibitem{Gadde:2009dj}
A.~Gadde, E.~Pomoni and L.~Rastelli,
\textit{``{The Veneziano Limit of N = 2 Superconformal QCD: Towards the String
  Dual of N = 2 SU(N(c)) SYM with N(f) = 2 N(c)}''},
\texttt{\arxivref{0912.4918}{arXiv:0912.4918}}.

\bibitem{Gadde:2010zi}
A.~Gadde, E.~Pomoni and L.~Rastelli,
\textit{``{Spin Chains in $\mathcal{N}$=2 Superconformal Theories: From the
  $\mathbb{Z}_{2}$ Quiver to Superconformal QCD}''},
\textsf{\doiref{10.1007/JHEP06(2012)107}{JHEP~1206,~107~(2012)}},
\texttt{\arxivref{1006.0015}{arXiv:1006.0015}}.

\bibitem{Pomoni:2011jj}
E.~Pomoni and C.~Sieg,
\textit{``{From N=4 gauge theory to N=2 conformal QCD: three-loop mixing of
  scalar composite operators}''},
\texttt{\arxivref{1105.3487}{arXiv:1105.3487}}.

\bibitem{Gadde:2012rv}
A.~Gadde, P.~Liendo, L.~Rastelli and W.~Yan,
\textit{``{On the Integrability of Planar $N=2$ Superconformal Gauge
  Theories}''},
\textsf{\doiref{10.1007/JHEP08(2013)015}{JHEP~1308,~015~(2013)}},
\texttt{\arxivref{1211.0271}{arXiv:1211.0271}}.

\bibitem{Pomoni:2013poa}
E.~Pomoni,
\textit{``{Integrability in N=2 superconformal gauge theories}''},
\textsf{\doiref{10.1016/j.nuclphysb.2015.01.006}{Nucl.~Phys.~B893,~21~(2015)}},
\texttt{\arxivref{1310.5709}{arXiv:1310.5709}}.

\bibitem{Mitev:2014yba}
V.~Mitev and E.~Pomoni,
\textit{``{Exact effective couplings of four dimensional gauge theories with
  $\mathcal N=$ 2 supersymmetry}''},
\textsf{\doiref{10.1103/PhysRevD.92.125034}{Phys.~Rev.~D92,~125034~(2015)}},
\texttt{\arxivref{1406.3629}{arXiv:1406.3629}}.

\bibitem{Mitev:2015oty}
V.~Mitev and E.~Pomoni,
\textit{``{Exact Bremsstrahlung and Effective Couplings}''},
\textsf{\doiref{10.1007/JHEP06(2016)078}{JHEP~1606,~078~(2016)}},
\texttt{\arxivref{1511.02217}{arXiv:1511.02217}}.

\bibitem{Pomoni:2019oib}
E.~Pomoni,
\textit{``{4D $\mathcal{N}=2$ SCFTs and spin chains}''},
\textsf{\doiref{10.1088/1751-8121/ab7f66}{J.~Phys.~A~53,~283005~(2020)}},
\texttt{\arxivref{1912.00870}{arXiv:1912.00870}}.

\bibitem{Pittelli:2019ceq}
A.~Pittelli and M.~Preti,
\textit{``{Integrable fishnet from $\gamma$-deformed $\mathcal{N}=2$
  quivers}''},
\textsf{\doiref{10.1016/j.physletb.2019.134971}{Phys.~Lett.~B~798,~134971~(2019)}},
\texttt{\arxivref{1906.03680}{arXiv:1906.03680}}.

\bibitem{Niarchos:2019onf}
V.~Niarchos, C.~Papageorgakis and E.~Pomoni,
\textit{``{Type-B Anomaly Matching and the 6D (2,0) Theory}''},
\textsf{\doiref{10.1007/JHEP04(2020)048}{JHEP~2004,~048~(2020)}},
\texttt{\arxivref{1911.05827}{arXiv:1911.05827}}.

\bibitem{Niarchos:2020nxk}
V.~Niarchos, C.~Papageorgakis, A.~Pini and E.~Pomoni,
\textit{``{(Mis-)Matching Type-B Anomalies on the Higgs Branch}''},
\textsf{\doiref{10.1007/JHEP01(2021)106}{JHEP~2101,~106~(2021)}},
\texttt{\arxivref{2009.08375}{arXiv:2009.08375}}.

\bibitem{Fiol:2020ojn}
B.~Fiol, J.~Martfnez-Montoya and A.~Rios~Fukelman,
\textit{``{The planar limit of $ \mathcal{N} $ = 2 superconformal quiver
  theories}''},
\textsf{\doiref{10.1007/JHEP08(2020)161}{JHEP~2008,~161~(2020)}},
\texttt{\arxivref{2006.06379}{arXiv:2006.06379}}.

\bibitem{Zarembo:2020tpf}
K.~Zarembo,
\textit{``{Quiver CFT at strong coupling}''},
\textsf{\doiref{10.1007/JHEP06(2020)055}{JHEP~2006,~055~(2020)}},
\texttt{\arxivref{2003.00993}{arXiv:2003.00993}}.

\bibitem{Ouyang:2020hwd}
H.~Ouyang,
\textit{``{Wilson loops in circular quiver SCFTs at strong coupling}''},
\textsf{\doiref{10.1007/JHEP02(2021)178}{JHEP~2102,~178~(2021)}},
\texttt{\arxivref{2011.03531}{arXiv:2011.03531}}.

\bibitem{Beccaria:2021ksw}
M.~Beccaria and A.~A.~Tseytlin,
\textit{``{$1/N$ expansion of circular Wilson loop in $\mathcal N=2$
  superconformal $SU(N)\times SU(N)$ quiver}''},
\textsf{\doiref{10.1007/JHEP04(2021)265}{JHEP~2104,~265~(2021)}},
\texttt{\arxivref{2102.07696}{arXiv:2102.07696}}.

\bibitem{Pini:2017ouj}
A.~Pini, D.~Rodriguez-Gomez and J.~G.~Russo,
\textit{``{Large $N$ correlation functions $ \mathcal{N}=$ 2 superconformal
  quivers}''},
\textsf{\doiref{10.1007/JHEP08(2017)066}{JHEP~1708,~066~(2017)}},
\texttt{\arxivref{1701.02315}{arXiv:1701.02315}}.

\bibitem{Billo:2022gmq}
M.~Bill\`o, M.~Frau, A.~Lerda, A.~Pini and P.~Vallarino,
\textit{``{Structure Constants in N=2 Superconformal Quiver Theories at Strong
  Coupling and Holography}''},
\textsf{\doiref{10.1103/PhysRevLett.129.031602}{Phys.~Rev.~Lett.~129,~031602~(2022)}},
\texttt{\arxivref{2206.13582}{arXiv:2206.13582}}.

\bibitem{Billo:2022fnb}
M.~Billo, M.~Frau, A.~Lerda, A.~Pini and P.~Vallarino,
\textit{``{Localization vs holography in 4d$ \mathcal{N} $ = 2 quiver
  theories}''},
\textsf{\doiref{10.1007/JHEP10(2022)020}{JHEP~2210,~020~(2022)}},
\texttt{\arxivref{2207.08846}{arXiv:2207.08846}}.

\bibitem{Gaberdiel:2022iot}
M.~R.~Gaberdiel and F.~Galvagno,
\textit{``{Worldsheet dual of free $ \mathcal{N} $ = 2 quiver gauge
  theories}''},
\textsf{\doiref{10.1007/JHEP10(2022)077}{JHEP~2210,~077~(2022)}},
\texttt{\arxivref{2206.08795}{arXiv:2206.08795}}.

\bibitem{Rey:2010ry}
S.-J.~Rey and T.~Suyama,
\textit{``{Exact Results and Holography of Wilson Loops in N=2 Superconformal
  (Quiver) Gauge Theories}''},
\textsf{\doiref{10.1007/JHEP01(2011)136}{JHEP~1101,~136~(2011)}},
\texttt{\arxivref{1001.0016}{arXiv:1001.0016}}.

\bibitem{Billo:2017glv}
M.~Billo, F.~Fucito, A.~Lerda, J.~F.~Morales, {\relax Ya}.~S.~Stanev and
  C.~Wen,
\textit{``{Two-point Correlators in N=2 Gauge Theories}''},
\textsf{\doiref{10.1016/j.nuclphysb.2017.11.003}{Nucl.~Phys.~B926,~427~(2018)}},
\texttt{\arxivref{1705.02909}{arXiv:1705.02909}}.

\bibitem{Gerchkovitz:2016gxx}
E.~Gerchkovitz, J.~Gomis, N.~Ishtiaque, A.~Karasik, Z.~Komargodski and
  S.~S.~Pufu,
\textit{``{Correlation Functions of Coulomb Branch Operators}''},
\textsf{\doiref{10.1007/JHEP01(2017)103}{JHEP~1701,~103~(2017)}},
\texttt{\arxivref{1602.05971}{arXiv:1602.05971}}.

\bibitem{Rodriguez-Gomez:2016ijh}
D.~Rodriguez-Gomez and J.~G.~Russo,
\textit{``{Large N Correlation Functions in Superconformal Field Theories}''},
\textsf{\doiref{10.1007/JHEP06(2016)109}{JHEP~1606,~109~(2016)}},
\texttt{\arxivref{1604.07416}{arXiv:1604.07416}}.

\bibitem{Kawamoto:2008gp}
S.~Kawamoto, T.~Kuroki and A.~Miwa,
\textit{``{Boundary condition for D-brane from Wilson loop, and gravitational
  interpretation of eigenvalue in matrix model in AdS/CFT correspondence}''},
\textsf{\doiref{10.1103/PhysRevD.79.126010}{Phys.~Rev.~D~79,~126010~(2009)}},
\texttt{\arxivref{0812.4229}{arXiv:0812.4229}}.

\bibitem{Preti:2017fjb}
M.~Preti,
\textit{``{WiLE: a Mathematica package for weak coupling expansion of Wilson
  loops in ABJ(M) theory}''},
\textsf{\doiref{10.1016/j.cpc.2017.12.011}{Comput.~Phys.~Commun.~227,~126~(2018)}},
\texttt{\arxivref{1707.08108}{arXiv:1707.08108}}.

\bibitem{Preti:2018vog}
M.~Preti,
\textit{``{STR: a Mathematica package for the method of uniqueness}''},
\textsf{\doiref{10.1142/S0129183120501466}{Int.~J.~Mod.~Phys.~C~31,~2050146~(2020)}},
\texttt{\arxivref{1811.04935}{arXiv:1811.04935}}.

\bibitem{Preti:2019rcq}
M.~Preti,
\textit{``{The Game of Triangles}''},
\textsf{\doiref{10.1088/1742-6596/1525/1/012015}{J.~Phys.~Conf.~Ser.~1525,~012015~(2020)}},
\texttt{\arxivref{1905.07380}{arXiv:1905.07380}}.

\bibitem{Gromov:2012eu}
N.~Gromov and A.~Sever,
\textit{``{Analytic Solution of Bremsstrahlung TBA}''},
\textsf{\doiref{10.1007/JHEP11(2012)075}{JHEP~1211,~075~(2012)}},
\texttt{\arxivref{1207.5489}{arXiv:1207.5489}}.

\bibitem{Gromov:2013qga}
N.~Gromov, F.~Levkovich-Maslyuk and G.~Sizov,
\textit{``{Analytic Solution of Bremsstrahlung TBA II: Turning on the Sphere
  Angle}''},
\textsf{\doiref{10.1007/JHEP10(2013)036}{JHEP~1310,~036~(2013)}},
\texttt{\arxivref{1305.1944}{arXiv:1305.1944}}.

\bibitem{Sizov:2013joa}
G.~Sizov and S.~Valatka,
\textit{``{Algebraic Curve for a Cusped Wilson Line}''},
\textsf{\doiref{10.1007/JHEP05(2014)149}{JHEP~1405,~149~(2014)}},
\texttt{\arxivref{1306.2527}{arXiv:1306.2527}}.

\bibitem{Fiol:2012sg}
B.~Fiol, B.~Garolera and A.~Lewkowycz,
\textit{``{Exact results for static and radiative fields of a quark in N=4
  super Yang-Mills}''},
\textsf{\doiref{10.1007/JHEP05(2012)093}{JHEP~1205,~093~(2012)}},
\texttt{\arxivref{1202.5292}{arXiv:1202.5292}}.

\end{thebibliography}

\end{document}